\documentclass{article}

\usepackage{PRIMEarxiv}

\usepackage[utf8]{inputenc} 
\usepackage[T1]{fontenc}    
\usepackage{hyperref}       
\usepackage{url}            
\usepackage{booktabs}       
\usepackage{amsfonts}       
\usepackage{nicefrac}       
\usepackage{microtype}      
\usepackage{lipsum}
\usepackage{fancyhdr}       
\usepackage{graphicx}       
\graphicspath{{media/}}     

\usepackage{algorithm}
\usepackage{algorithmic}
\usepackage{multirow}
\usepackage{subcaption}
\usepackage{amsmath}

\title{Studying the Effect of Audio Filters in Pre-Trained Models for Environmental Sound Classification

}

\author{
  Aditya Dawn \\
  University of Kalyani \\
  Kalyani, India\\
  \texttt{adityadawn98@gmail.com} \\
   \And
  Wazib Ansar \\
  A.K.Choudhury School of IT \\
  University of Calcutta \\
  Kolkata, India\\
  \texttt{waakcs\_rs@caluniv.ac.in} \\
}

\begin{document}
\maketitle

\begin{abstract}
Environmental Sound Classification is an important problem of sound recognition and is more complicated than speech recognition problems as environmental sounds are not well structured with respect to time and frequency. Researchers have used various CNN models to learn audio features from different audio features like log mel spectrograms, gammatone spectral coefficients, mel-frequency spectral coefficients, generated from the audio files, over the past years. In this paper, we propose a new methodology : Two-Level Classification; the Level 1 Classifier will be responsible to classify the audio signal into a broader class and the Level 2 Classifiers will be responsible to find the actual class to which the audio belongs, based on the output of the Level 1 Classifier. We have also shown the effects of different audio filters, among which a new method of Audio Crop is introduced in this paper, which gave the highest accuracies in most of the cases. We have used the ESC-50 dataset for our experiment and obtained a maximum accuracy of 78.75\% in case of  Level 1 Classification and 98.04\% in case of Level 2 Classifications.
\end{abstract}


\section{Introduction}
Environmental Sound Classification (ESC) has become a challenging job in recent times. The classification and identification of environmental sounds, which include dog barking, birds chirping, knocking on the door, the sound of a vacuum cleaner, car horn, water drops and many other similar sounds, are necessary for developing smart-home appliances \cite{smarthome}, security systems \cite{security, machinehearing}, etc. to make human life more secure. For example, if mobile devices are able to recognize the sound of car honking then road accidents and pedestrian accidents will decrease by a significant rate. Or, if the auto-pet care systems can identify the sounds of dogs or birds, they can provide the animals with food and water. Or if a smart security system can recognize door knocking sounds it will be able to notify the owner of the place about the presence of a person behind the door. In other words, a well-featured and trained sound classifier will be able to recognize the sound from human surroundings or environment and will be able to make human life safe and sound.

Traditional Sound Processing is based on the sound features like log mel spectrograms with delta informations \cite{piczak2015cnn}, Gammatones \cite{agrawal2017} and Mel Frequency Cepstral Coefficients (MFCC) \cite{baritelli2008mfcc}. Various Machine Learning and Deep Learning techniques are applied with these features to obtain high scoring results. Though the Machine Learning algorithms were able to obtain good results, in recent years the breakthrough of Deep Neural Networks, mainly Convolutional Neural Networks has been significant. CNN's were used with the sound features to obtain accuracies like 83.9\% \cite{zhichao2018}, 86.95\% \cite{zhichao2019} to high as 97.57\% \cite{mushtaq2021}.

In this paper, we propose a two-level classification method using CNN models of the classes VGG, ResNet and EfficientNet with audio modifiers like PCEN, Spectral Gating(Noise Removal), Audio Crop and Audio Filters like Low Pass Filter, High Pass Filter, Band Pass Filter and Band Stop Filter. The Level 1 classifier will classify the sounds into broader groups of Animals, Birds, Nature, etc., while the Level 2 classification will pull the audio signal to the sub-class. For example, if an audio signal is classified to the Animal class by the Level 1 classifier, then the Level 2 classifier will be responsible to detect the actual animal, which might be a dog or a cow or a sheep. After applying the CNN models and the audio modifiers, our method obtained a score of 78.75\% in the case of the Level 1 Classification and the maximum score obtained in the case of the Level 2 Classifiers is 98.04\%. 

The remaining portion of this article is arranged descriptively and divided into 7 sections. Sound Classification processes used in previous works for the similar problem are explained in Section \ref{related works}. In Section \ref{methods}, we have discussed CNN and Audio Modifiers which were used for our work. In Section \ref{proposed method}, we have shown the process of our method with a flowchart, with a suitable description. The tools and libraries are discussed in Section \ref{implementation}. Then, the results of our works are shown in Section \ref{results} and discussed in Section \ref{discussion}. Finally, we conclude our paper in Section \ref{conclusion}.

\section{Related Works}
\label{related works}

In this section, we discuss the previous works by researchers on the ESC problem.

Piczak extracted the log mel spectrograms for each frame of the audio files. Piczak used these log mel spectrograms and their delta informations in the CNN model proposed by him in \cite{piczak2015cnn}. He got an accuracy of 64.5\% with this approach. The goal of Piczak's paper was to evaluate the success of CNNs, when applied to ESC tasks.

Agrawal et al. \cite{agrawal2017}, used a TEO-based gammatone feature set for the problem. Firstly, they extracted the gammatone filterbanks from the raw audio files and applied a bandpass filter. Then, they applied a half-wave rectifier on each of the sub-bands and then Teager Energy Operator was applied again on each of the sub-bands. Finally, they applied short-term averaging and short-term spectral features were obtained. The obtained spectrograms were given as input to a CNN architecture similar to that used by Piczak \cite{piczak2015cnn} and obtained a score of 81.95\%. However, they showed that the TEO-based Gammatone Spectral Coefficients failed to give better results with CNN.

Zhang et al. \cite{xiaohu2017} showed the performance of Dilated Convolution Network with seven layers and two input channels. They used log mel spectrograms and delta feature spectrograms in the proposed Dilated Convolution Network. They also studied the classification accuracy obtained for ReLU-type activation functions. However, they got a classification accuracy of 68.1\% on the ESC-50 dataset and it was also noted by them that the improvement of classification accuracy cost more of higher computational complexity and bigger storage.

Zhichao et al. \cite{zhichao2018} proposed a new CNN architecture inspired by VGG Net by using 1-D convolution filters in place of $ 3 \times 3 $ convolution filters, to learn local patterns across frequency and time. They extracted log mel spectrograms and gammatone spectrograms, which were used in the proposed CNN architecture, along with their delta information and achieved a classification accuracy of 83.9\%. 

In \cite{zhichao2019}, Zhichao et al. adopted a convolutional RNN architecture for the problem.At first, they used CNN with channel temporal attention mechanism in convolution layers, with log-gammatone spectrograms to extract high-level features from the spectrograms, which were further used in the bidirectional gated recurrent unit to analyse temporal correlations. They showed a classification accuracy of 86.5\% on the ESC-50 dataset.

In \cite{ullo2020}, Ullo et al. proposed a method that uses a hybrid structure made of OAS, STFT, CNN and different classification techniques for the classification of the classes in the ESC-10 dataset and they achieved a classification accuracy of 95.8\%.

Mushtaq et al. \cite{mushtaq2021} showed the performances of different data augmentations on the audio files and log mel spectrograms of the original audio files and augmented audio files with transfer learning and obtained an accuracy of 97.57\% for ESC-50 dataset. They also showed the performances of distinct pre-trained models, which included ResNet, DenseNet, AlexNet, SqueezeNet and VGG. 

Ansar et al. \cite{ansar2024efficientnet} proposed an EfficientNet ensemble with triple-layered approach to eliminate noise for classification of audio signals. They further validated a trade-off between model depth and number of parameters to obtain optimal accuracy through extensive evaluation on a bouquet of models.

It is clear from the previous works that, all the new methodologies and new models were built to classify the audio files to 50 classes directly. So in this study, a new approach is proposed, a two-level classification,  with the spectrograms obtained after the application of audio modifiers and the pre-trained CNN models, which obtained a state of art accuracy score compared to the works discussed above.

\section{Methods}
\label{methods}

In this section, we discuss about the types of CNNs and audio modifiers used for our project.

\subsection{Convolutional Neural Network (CNN)}
\label{subsec:CNN}

\begin{figure}[htp]
    \centering
    \includegraphics[width=\textwidth]{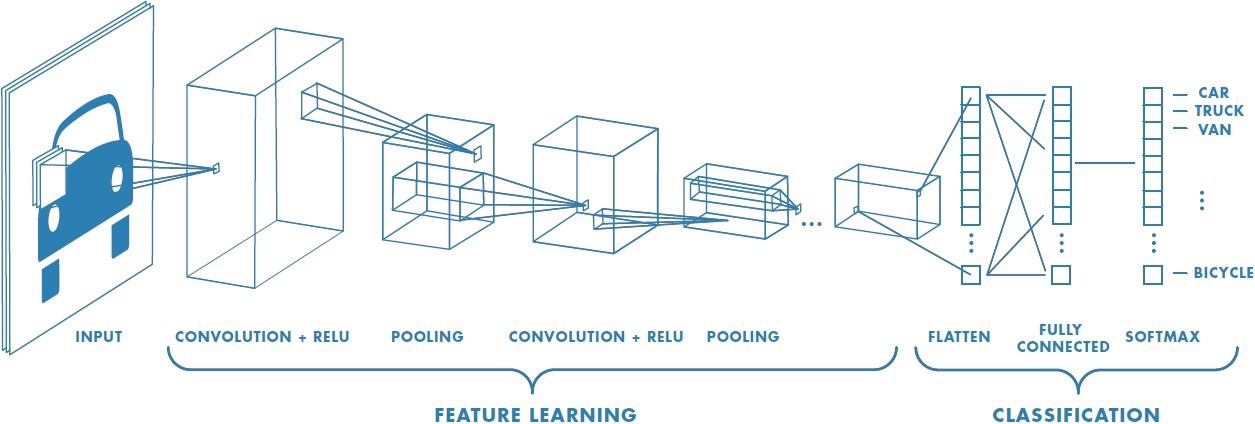}
    \caption{CNN architecture}
    \label{fig:cnn}
\end{figure}

CNNs are used in the application of image processing. CNNs can be trained well to understand the hidden features of the images. This is because CNN applies different relevant filters and the architecture reduces the number of parameters involved and increases the re-usability of the weights. For this reason, the spatial and temporal dependencies of an image are successfully captured by CNN. CNNs are mostly used for classification and computer vision tasks. Fig \ref{fig:cnn}\footnote{\href{https://towardsdatascience.com/a-comprehensive-guide-to-convolutional-neural-networks-the-eli5-way-3bd2b1164a53}{\nolinkurl{towardsdatascience}}} shows a general architecture of a CNN Model.

\vspace{5mm}

\vspace{5mm}

CNN models consists mainly of three layers :

\begin{enumerate}
    \item Convolution Layer
    \item Pooling Layer
    \item Fully-connected Layer
\end{enumerate}

LeCun et al. \cite{lecunn1990} introduced the first CNN architecture, LeNet-5 for the recognition of handwritten digits (input images were of dimension $32 \times 32 \times 1$), using the MNIST dataset \footnote{\href{http://yann.lecun.com/exdb/mnist/}{\nolinkurl{MNIST database}}}. LeNet-5 was a vary shallow CNN with alternating convolution layers and pooling layers and had only about 60,000 parameters.

AlexNet was then introduced by Krizhevsky et al. \cite{krizhevsky2012}. The network was similar to LeNet but instead of alternating convolution layers and pooling layers, AlexNet had all the convolution layers stacked together. Also compared to LeNet-5, this network is much bigger and deeper.


Later, VGGNet was introduced by Simonyan \& Zisserman \cite{vgg16}. Earlier, models like AlexNet used high dimensional filter in the initial layers, but VGG changed this by using $3 \times 3$ filters.

He et al. \cite{resnet} have presented a residual learning framework where the layers learn residual functions with respect to the inputs received instead of learning un-referenced functions. They were able to prove that this work is particularly useful for training deeper networks since residual networks are easier to optimize and gain much accuracy. The main drawback of this network is that it is much expensive to evaluate due to the huge number of parameters.

Various models were thus developed focusing either on performance or computational efficiency. Tan \& V.Le introduced EfficientNet \cite{efficientnet} model, which was able to solve both the problems. They proposed a common CNN architecture, which worked with three parameters width, depth and resolution. Width refers to the number of channels present in various layers, depth refers to the number of layers in the model and resolution refers to the input image size for the model. EfficientNet mainly helps in performing compound scaling with depth, width and resolution of the image. The compound scaling method only enhances the predictive capacity of the networks by replicating base network’s underlying convolutional operations and network structure.

In this paper we have also shown the efficiency of 3 classes of pre-trained models, which are as follows :

\begin{itemize}
    \item VGG \cite{vgg16}
        \begin{itemize}
            \item VGG16
            \item VGG19
        \end{itemize}
    \item ResNet \cite{resnet}
        \begin{itemize}
            \item ResNet50
            \item ResNet101
            \item ResNet152
        \end{itemize}
    \item EfficientNet \cite{efficientnet}
        \begin{itemize}
            \item EfficientNetB0
            \item EfficientNetB1
            \item EfficientNetB2
            \item EfficientNetB3
            \item EfficientNetB4
        \end{itemize}
\end{itemize}

\subsection{Spectrogram}

A Spectrogram, usually depicted as a heatmap, is a visual representation, of a spectrum of frequencies of a signal as it varies with time. Spectrograms of some audio files of each category of class based on Table \ref{tab:new-classification} are shown in Fig \ref{fig:spectrograms}. The spectrograms are passed into CNN Models to learn the audio features.

\subsection{Understanding Audio Modifiers Used}

\subsubsection{Spectral Gating}

Spectral Gating as explained by \cite{spectralgating} is a technique, which is comprised of several steps. A Fourier transformation is applied on the noise-only portion of the audio signal to create a spectral "fingerprint", which is further used as a "gate" to filter the audio signal. The frequencies in the audio signals, which are above the gated value are passed, while those below the value are removed. Here, we have removed noise based on the principle of Spectral Gating. The result of Spectral Gating performed on an audio file is shown in Fig \ref{fig:sg} with the audio form and spectrogram.

\subsubsection{Per Channel Energy Normalization (PCEN)}

\cite{pcen1} introduced PCEN as an alternative to the log-mel frontend. \cite{pcen2} discussed the working principle of PCEN. PCEN is the result of three-component operation :

\begin{enumerate}
    \item Temporal integration 
    \item Adaptive Gain Control
    \item Dynamic Range Compression
\end{enumerate} 

The result of PCEN on two audio files is shown in Fig \ref{fig:pcen}.

\subsubsection{Audio Crop}

The main idea behind introducing this feature in our work, is to repeat the non-zero portions of an audio sample over the maximum time length of the audio files provided with the data. This method can be explained better with algorithms \ref{audiocrop1} and \ref{audiocrop2}. Algorithm \ref{audiocrop1} finds the maximum time length present among the audio files and algorithm \ref{audiocrop2} removes the silent portions of the audio files. The process of Audio Cropping is shown in Fig \ref{fig:aud_crop}.

\subsubsection{Audio Filters}

Research and developments on Audio Filters have been done for audio modifications as mentioned in  \cite{audiofilters}. The filters that we have used in our work modifies the audio signals based on frequencies.

\begin{enumerate}

    \item \textbf{Low Pass Filter} : Low Pass Filter allows the frequencies lower than a cut-off frequency to pass and attenuates the frequencies higher than the cut-off frequency.
    
    \item \textbf{High Pass Filter} : High Pass Filter allows the frequencies higher than a cut-off frequency to pass and attenuates the frequencies lower than the cut-off frequency.
    
    \item \textbf{Band Pass Filter} : Band Pass Filter accepts two cut-off frequencies, low-cut frequency and high-cut frequency. This Filter allows the band of frequencies within the low-cut and high-cut frequencies to pass and attenuates the frequencies lower than the low-cut frequency and higher than the high-cut frequency.

    \item \textbf{Band Stop Filter} : Band Stop Filter also accepts two cut-off frequencies, low-cut frequency and high-cut frequency, like Band Pass Filter. But like Band Pass Filter, this Filter attenuates the band of frequencies within the low-cut and high-cut frequencies and allows the frequencies lower than the low-cut frequency and higher than the high-cut frequency to pass.
    
\end{enumerate}

\section{Proposed Method}
\label{proposed method}

\begin{figure}[h]
    \centering
    \includegraphics[width=\textwidth]{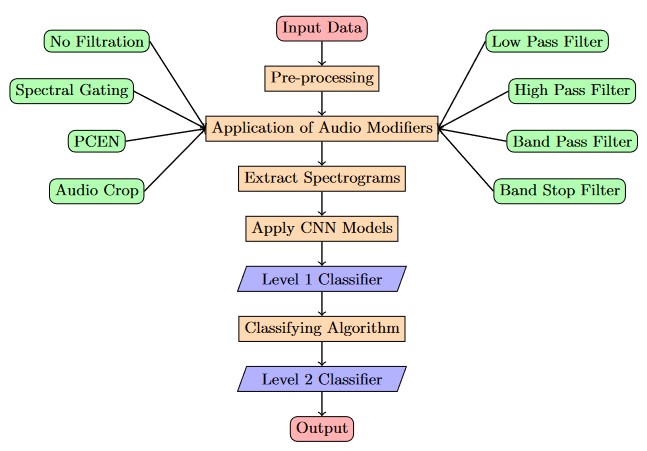}
    \caption{Illustration of the proposed workflow}
    \label{fig:wf}
    \vspace{1pt}
\end{figure}

In this section, we discuss the design of the Two-Level Classification method, which we are proposing in this
paper, as shown in Figure \ref{fig:wf}.

\vspace{15pt}

\vspace{5pt}

\begin{table}[h]
\centering
\caption{New Classification}
\label{tab:new-classification}
\resizebox{\textwidth}{!}{%
\begin{tabular}{ccc|c|cc|c|}
\hline
\multicolumn{1}{|c|}{\textbf{Animal}} & \multicolumn{1}{c|}{\textbf{Birds}} & \textbf{\begin{tabular}[c]{@{}c@{}}Natural\\ Soundscapes\end{tabular}} & \textbf{Human}    & \multicolumn{1}{c|}{\textbf{\begin{tabular}[c]{@{}c@{}}Machine\\ Sounds\end{tabular}}} & \textbf{\begin{tabular}[c]{@{}c@{}}Domestic\\ sounds\end{tabular}} & \textbf{\begin{tabular}[c]{@{}c@{}}Outdoor\\ noises\end{tabular}} \\ \hline
\multicolumn{1}{|c|}{Dog}             & \multicolumn{1}{c|}{Chirping Birds} & Rain                                                                   & Crying Baby       & \multicolumn{1}{c|}{Mouse Click}                                                       & Door knock                                                         & Helicopter                                                        \\ \hline
\multicolumn{1}{|c|}{Sheep}           & \multicolumn{1}{c|}{Rooster}        & Sea Waves                                                              & Sneezing          & \multicolumn{1}{c|}{Keyboard Typing}                                                   & Toilet flush                                                       & Chainsaw                                                          \\ \hline
\multicolumn{1}{|c|}{Pig}             & \multicolumn{1}{c|}{Crow}           & Crackling Fire                                                         & Clapping          & \multicolumn{1}{c|}{Washing Machine}                                                   & Clock alarm                                                        & Siren                                                             \\ \hline
\multicolumn{1}{|c|}{Cow}             & \multicolumn{1}{c|}{Hen}            & Wind                                                                   & Breathing         & \multicolumn{1}{c|}{Vacuum cleaner}                                                    & Door, wood creaks                                                  & Car Horn                                                          \\ \hline
\multicolumn{1}{|c|}{Frog}            & \multicolumn{1}{c|}{}               & Pouring water                                                          & Coughing          & \multicolumn{1}{c|}{}                                                                  & Can opening                                                        & Engine                                                            \\ \cline{1-1} \cline{3-4} \cline{6-7} 
\multicolumn{1}{|c|}{Cat}             & \multicolumn{1}{c|}{}               & Water drops                                                            & Footsteps         & \multicolumn{1}{c|}{}                                                                  & Clock tick                                                         & Train                                                             \\ \cline{1-1} \cline{3-4} \cline{6-7} 
\multicolumn{1}{|c|}{Insects(flying)} & \multicolumn{1}{c|}{}               & Thunderstorm                                                           & Laughing          & \multicolumn{1}{c|}{}                                                                  & Glass breaking                                                     & Church bells                                                      \\ \cline{1-1} \cline{3-4} \cline{6-7} 
\multicolumn{1}{|c|}{Crickets}        &                                     &                                                                        & Brushing teeth    &                                                                                        &                                                                    & Airplane                                                          \\ \cline{1-1} \cline{4-4} \cline{7-7} 
                                      &                                     &                                                                        & Snoring           &                                                                                        &                                                                    & Fireworks                                                         \\ \cline{4-4} \cline{7-7} 
                                      &                                     &                                                                        & Drinking, sipping &                                                                                        &                                                                    & Hand saw                                                          \\ \cline{4-4} \cline{7-7} 
\end{tabular}%
}
\vspace{7mm}
\end{table}

\begin{enumerate}
    \item [Step 1] The audio files of ESC-50 dataset is taken as input.
    
    \item [Step 2] In the pre-processing part, for our work, we have divided the dataset into 7 groups as shown in table \ref{tab:new-classification}. This new divisions were made based on the origin or source of the sounds. For example, Dog, Sheep, Pig, etc are evidently animals. So, they are grouped in the "Animal" class. Rooster, Crow, Hen and Chirping Birds are placed in the "Bird" class. Rain, Sea Waves, Wind, Pouring Water, Thunderstorms are observed in nature and so they are grouped in the "Natural Soundscapes" class. Sneezing, Clapping, Breathing, Drinking, etc are parts of human behavior. So, they are placed in the "Human" group.
    Mouse Click, Keyboard Typing, Washing Machine and Vacuum Cleaner are sound originating from some particular machines. So, they are grouped in the "Machine Sounds" class. Door knock, Toilet flush, Clock alarm, Can opening, etc are found in domestic environment. So, they are grouped in the "Domestic sounds" class. Finally, as Helicopter, Chainsaw, Siren, etc are found in outdoor spaces and so they have been grouped in the "Outdoor noises" class.

    \item [Step 3] Audio modifiers were then applied to the audio files. The audio modifiers include :
        \begin{enumerate}
            \item Spectral Gating
            \item PCEN
            \item Audio Crop
            \item Low Pass Filter
            \item High Pass Filter
            \item Band Pass Filter
            \item Band Stop Filter
        \end{enumerate}
        
    \item [Step 4] The Spectrograms of the modified audio files are extracted.
    
    \item [Step 5] Extracted Spectrograms are then passed to the CNN models mentioned in sub-section \ref{subsec:CNN} for the purpose of Level 1 Classification.
    
    \item [Step 6] The output from the Level 1 Classification is noted and used in algorithm \ref{classifying algo}.
    
    \item [Step 7] Algorithm \ref{classifying algo} will then identify the classifier in the Level 2 Classification stage to identify the actual class to which the audio file belongs.
    
\end{enumerate}

\vspace{.1mm}

Note that, we are using 10 CNN models as discussed in Section \ref{methods}, where we are passing eight different types of spectrogram images - one without filtration and seven with filtrations of Spectral Gating, PCEN, Audio Crop, Low Pass Filter, High Pass Filter, Band Pass Filter and Band Stop Filter. So for each classification, we are comparing the results of 80 different models.

\vspace{5mm}
\begin{algorithm}[H]

    \caption{Max Time Len}
    \label{audiocrop1}

    \begin{algorithmic}[1]
    
        \REQUIRE Path of the audio files in the data provided
        
        \ENSURE Time length of the longest audio file
    
        \STATE Time\_Lengths $\gets$ An empty list to store the lengths of the audio files
    
        \FORALL{Audio files}
    
        \STATE time\_length $\gets$ Length of the audio file
    
        \STATE Store time\_length in Time\_Lengths
    
        \ENDFOR
        
        \STATE Max\_time\_len $\gets$ Maximum value in Time\_Lengths
        
        \RETURN{} Max\_time\_len
    
    \end{algorithmic}

\end{algorithm}

\section{Implementation of the Method}
\label{implementation}

\subsection{Dataset : ESC-50}

\par We have used the ESC-50 \cite{piczak2015data} dataset for our work, which is a collection of 2000 recordings with an average duration of 5 seconds and a sampling frequency rate of 44100 Hz. These recordings have been collected from Freesound.org \footnote{\url{https://freesound.org/}}. The dataset consists of recordings of 40 audio files for each of the 50 categories. 

\vspace{-5pt}
\begin{algorithm}[H]

    \caption{Crop Audios according to the output of algorithm \ref{audiocrop1}}
    \label{audiocrop2}
    
    \begin{algorithmic}
    
    \REQUIRE
        \begin{enumerate}
            \item  Path of the Audio files in the data provided
            \item Max\_time $\gets$ Output of Algorithm \ref{audiocrop1}
            
        \end{enumerate}
    \ENSURE Audio files without silent portions
    
    \FORALL{Audio Files}
    
        \STATE time\_len $\gets$ time length of the audio file
        
        \STATE audio\_ $\gets$ audio[audio $\neq$ 0]
        
        \STATE audio\_add $\gets$ audio[audio $\neq$ 0]
        
        \STATE q $\gets$ quotient(Max\_time $\div$ time\_len)
        
        \IF{q $>$ 1}
            
            \STATE times $\gets$ q-1
            
            \FOR{time $\in$ times}
            
                \STATE audio\_ = concatenate(audio\_, audio\_add)
            
            \ENDFOR
        
        \ELSE
            
            \STATE pass
            
        \ENDIF
        
        \STATE r $\gets$ remainder(Max\_time $divide$ time\_len)
        
        \IF{r == 0}

            \STATE Pass

        \ELSE

            \STATE j $\gets$ 0

            \WHILE{j $\neq$ r}

                \STATE audio\_ = append(audio\_, audio\_[j]) 
                
            \ENDWHILE
            
        \ENDIF
        
    \ENDFOR
        
    \RETURN Cropped\_Audio $\gets$ audio\_
    
    \end{algorithmic}

\end{algorithm}

These 50 categories are generally grouped into 5 groups as done in the previous works mentioned in the Section \ref{related works}, which included the ESC-50 dataset. The groups are :

\begin{itemize}
    \item Animals
    \item Natural soundscapes \& water sounds
    \item Human, non-speech sounds
    \item Interior/domestic sounds
    \item Exterior/urban noises
\end{itemize}

\subsection{Libraries}

The work has been done using the Python Programming Language. The Python libraries used for this project are discussed below.

\subsubsection{NumPy}
The \texttt{NumPy} library \footnote{\url{https://numpy.org/}} helps in performing complex mathematical operations with arrays and random number generations. In this project we have used NumPy for random split of the dataset to make Training and Testing Sets to train and test on the CNN Models, respectively.

\subsubsection{Pandas}
The \texttt{Pandas} library \footnote{\url{https://pandas.pydata.org/}} helped in manipulating the dataframe and also performing some basic operations on the dataframe.

\vspace{5pt}
\begin{algorithm}[H]

    \caption{Classifying Algorithm}
    \label{classifying algo}
    
    \begin{algorithmic}[1]
    
        \REQUIRE Input $\gets$ Level 1 Classifier Output
        \ENSURE Output $\gets$ Actual\_Class
        
        \IF{Input == "Animal"}
            \STATE Actual\_Class = animal\_model.predict()
        
        \ELSIF{Input == "Bird"}
            \STATE Actual\_Class = bird\_model.predict()
            
        \ELSIF{Input == "Nature"}
            \STATE Actual\_Class = nature\_model.predict()
            
        \ELSIF{Input == "Human"}
            \STATE Actual\_Class = human\_model.predict()
            
        \ELSIF{Input == "Machine Sounds"}
            \STATE Actual\_Class = machine\_sounds\_model.predict()
            
        \ELSIF{Input == "Domestic"}
            \STATE Actual\_Class = domestic\_model.predict()
            
        \ELSE
            \STATE Actual\_Class = outdoor\_model.predict()
            
        \ENDIF

        \RETURN Actual\_Class

    \end{algorithmic}

    \vspace{5pt}

\end{algorithm}

\subsubsection{Matplotlib and Seaborn}
\texttt{Matplotlib} \footnote{\url{https://matplotlib.org/}} and \texttt{Seaborn} \footnote{\url{https://seaborn.pydata.org/}} were used for plottings.

\subsubsection{Librosa}
\texttt{Librosa} \footnote{https://librosa.org} was used to work with the audio files. Librosa helped in extracting the audio files from their respective locations. \texttt{Librosa} also has functions to extract mel spectrograms and functions for audio modifications which were used on the audio samples.

\subsubsection{SciPy}
\texttt{SciPy} \footnote{\url{https://scipy.org/}} is a library which has the functions to implement the audio filters on the audio samples provided in the data.

\subsubsection{Noise Reduce}
The \texttt{noisereduce}\footnote{\url{https://pypi.org/project/noisereduce/}} is used for Noise Removal from the audio files provided with the dataset.

\subsubsection{Tensorflow and Keras}
\texttt{Tensorflow}\footnote{\url{https://www.tensorflow.org/}} and \texttt{Keras}\footnote{\url{https://keras.io/}} were used to implement the CNN models. The pre-trained models from \texttt{Keras} were used in this work.

\subsubsection{Model Hyperparameters}

After the pre-trained layers of the pre-trained models, we added a layer with global average pooling method. After the global average layer, two dense layers were added each with 512 filters and activation function ReLU. The kernel initializer of the first dense layer was set to  glorot uniform. Stochastic gradient descent as the optimizer function during compilation of the models. 

Fig \ref{fig:spectrograms} shows that the intensity of sound is maximum within the range of 0 -- 512 Hz, while it decreases slightly to 2048 Hz and starts to fade after that in the cases of thunderstorm, vacuum cleaner, glass breaking, train sounds. Spectrograms of chirping birds, clapping show that the intensity is faded till 128 Hz. Again, in the cases of clapping and dog maximum intensity is visible mostly within 512 -- 4096 Hz. Besides this, the black portion indicates silence in the audio files of dog and glass breaking. Based o these observations, in case of the Audio Filters, we have used 512 Hz as the lower threshold and 2048 Hz as the higher threshold frequencies.

\begin{table}[H]
\centering
\caption{Cut-off Frequencies for Audio Filters}
\label{tab:cut-off-frequencies}
\begin{tabular}{|l|ll|}
\hline
\textbf{Audio Filter}                      & \multicolumn{2}{l|}{\textbf{Cut-off frequency}} \\ \hline
\textbf{Low Pass Filter}                   & \multicolumn{2}{l|}{512 Hz}                     \\ \hline
\textbf{High Pass Filter}                  & \multicolumn{2}{l|}{2048 Hz}                    \\ \hline
\multirow{2}{*}{\textbf{Band Pass Filter}} & \multicolumn{1}{l|}{lower cut-off}   & 512 Hz   \\ \cline{2-3} 
                                           & \multicolumn{1}{l|}{higher cut-off}  & 2048 Hz  \\ \hline
\multirow{2}{*}{\textbf{Band Stop Filter}} & \multicolumn{1}{l|}{lower cut-off}   & 512 Hz   \\ \cline{2-3} 
                                           & \multicolumn{1}{l|}{higher cut-off}  & 2048 Hz  \\ \hline
\end{tabular}%
\end{table}

\subsection{Implementation Details}

For our convenience, for each CNN Model, we first divided the data for the respective model in the ratio of 8:2 to create Training Set and the Testing Set. Then we divide the Training set again in the ratio of 8:2, as shown in table \ref{tab:number-of-samples}. Also, we have used a sampling rate of 44.1 KHz for the audio files.

\section{Results}
\label{results}

From table \ref{tab:number-of-samples}, it is clear that the number of testing samples is less except the case of Level 1 Classification. We are going to compare the performances of the classifiers and the models based on Classification Accuracy for Level 1 classifiers (table \ref{tab:level-1-classifier}) only, but in the other cases we are going to judge based on Highest Validation Accuracy obtained as a single miss-classification by the model will decrease the Classification Accuracy significantly, specifically in the cases of Birds and Machine Sounds classification, though we have provided both the Classification Accuracy and Validation Accuracy in the result tables of the classification models for Animals (table \ref{tab:animal-results}), Birds (table \ref{tab:bird-results}), Natural Soundscapes (table  \ref{tab:nature-results}), Human (table \ref{tab:human-results}), Machine Sounds (table \ref{tab:machine-sounds-results}), Domestic (table  \ref{tab:domestic-results}) and Outdoor noises (table \ref{tab:outdoor-results}).

\vspace{-5pt}
\begin{table}[h]
\centering
\caption{Distribution of Samples for Training, Validation and Testing}
\label{tab:number-of-samples}
\resizebox{\textwidth}{!}{%
\begin{tabular}{|c|c|c|c|c|c|}
\hline
\textbf{Mode of Classification} &
  \textbf{\begin{tabular}[c]{@{}c@{}}Number of \\ classes\end{tabular}} &
  \textbf{\begin{tabular}[c]{@{}c@{}}Total Number\\ of Samples\end{tabular}} &
  \textbf{\begin{tabular}[c]{@{}c@{}}Size of \\ Training Set\end{tabular}} &
  \textbf{\begin{tabular}[c]{@{}c@{}}Size of\\ Validation Set\end{tabular}} &
  \textbf{\begin{tabular}[c]{@{}c@{}}Size of\\ Testing Set\end{tabular}} \\ \hline
\textbf{Level 1 Classification} & 7  & 2000 & 1280 & 320 & 400 \\ \hline
\textbf{Animals}                & 8  & 320  & 205  & 51  & 64  \\ \hline
\textbf{Birds}                  & 4  & 160  & 103  & 25  & 32  \\ \hline
\textbf{Nature}                 & 7  & 280  & 180  & 44  & 56  \\ \hline
\textbf{Human}                  & 10 & 400  & 256  & 64  & 80  \\ \hline
\textbf{Machine Sounds}         & 4  & 160  & 103  & 25  & 32  \\ \hline
\textbf{Domestic}               & 8  & 320  & 205  & 51  & 64  \\ \hline
\textbf{Outdoor}                & 10 & 400  & 256  & 64  & 80  \\ \hline
\end{tabular}%
}
\vspace{5mm}
\end{table}
\vspace{-5pt}

\begin{figure}[h]
    \centering
    \includegraphics[width=\textwidth]{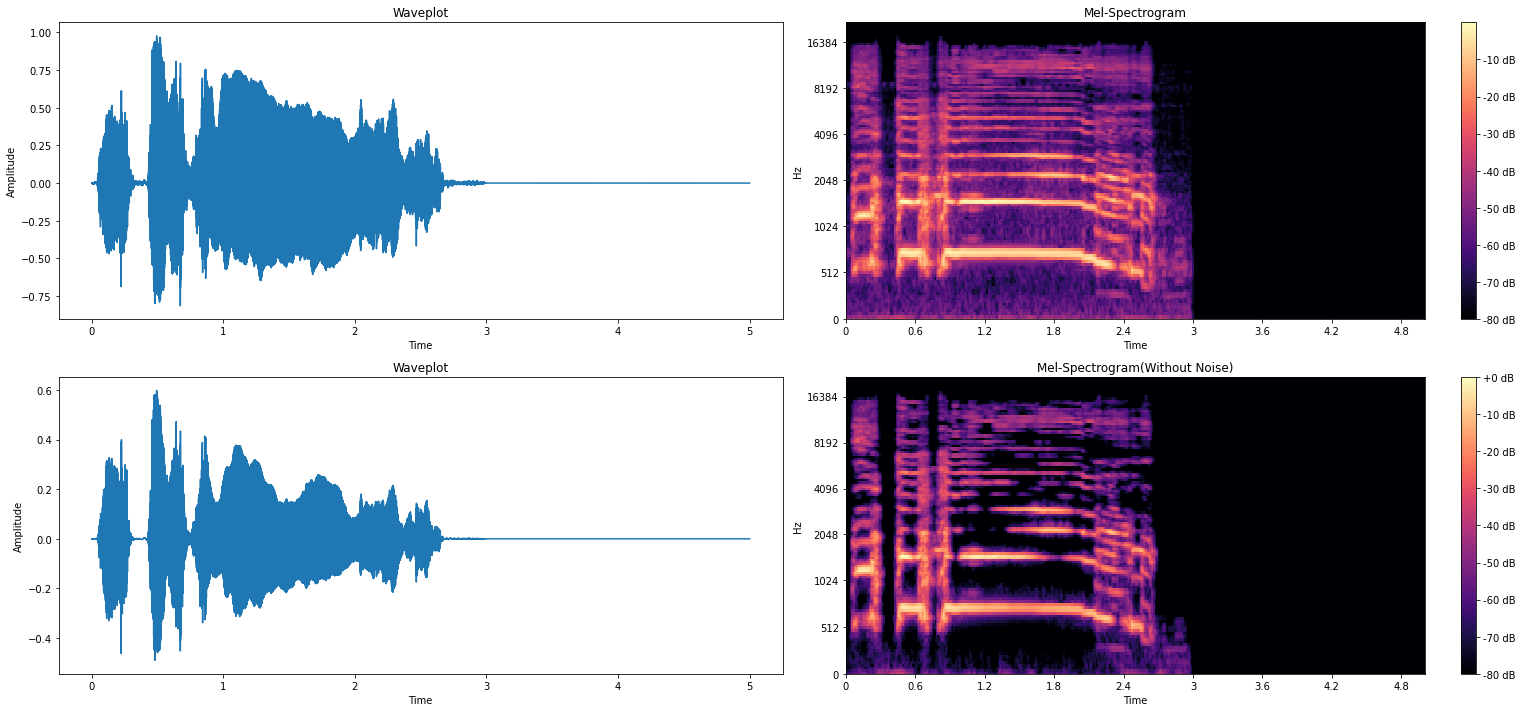}
    \caption{Effect of Spectral Gating}
    \label{fig:sg}
\end{figure}

\begin{figure}[H]
    \centering
    
    \begin{subfigure}{\textwidth}
    
        \includegraphics[width = \textwidth]{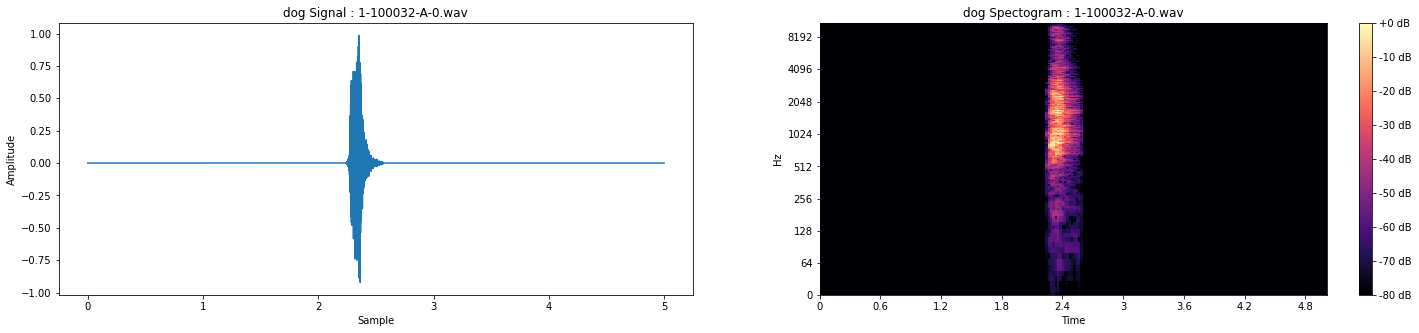}

        \caption{Spectrogram of an audio of dog -- barking}
    
    \end{subfigure}
    
    \begin{subfigure}{\textwidth}
    
        \includegraphics[width = \textwidth]{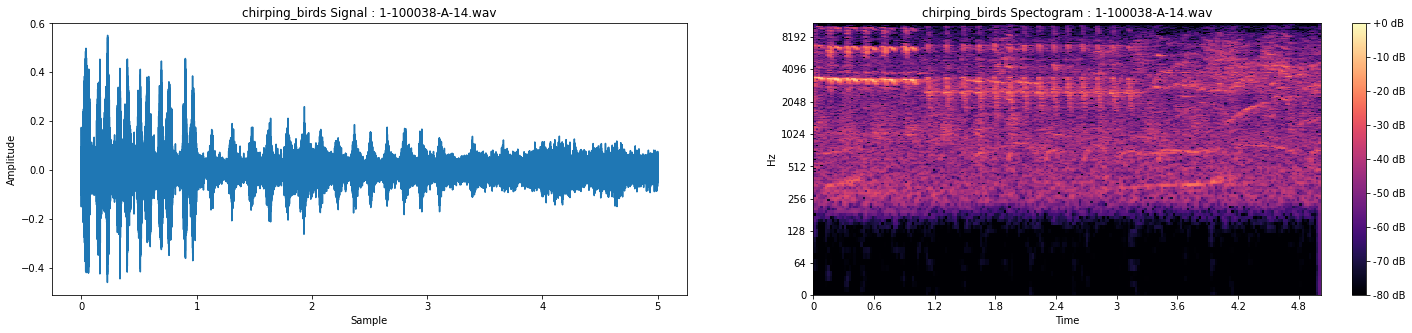}
        
        \caption{Spectrogram of an audio file of birds chirping}
    
    \end{subfigure}
    
    \begin{subfigure}{\textwidth}
    
        \includegraphics[width = \textwidth]{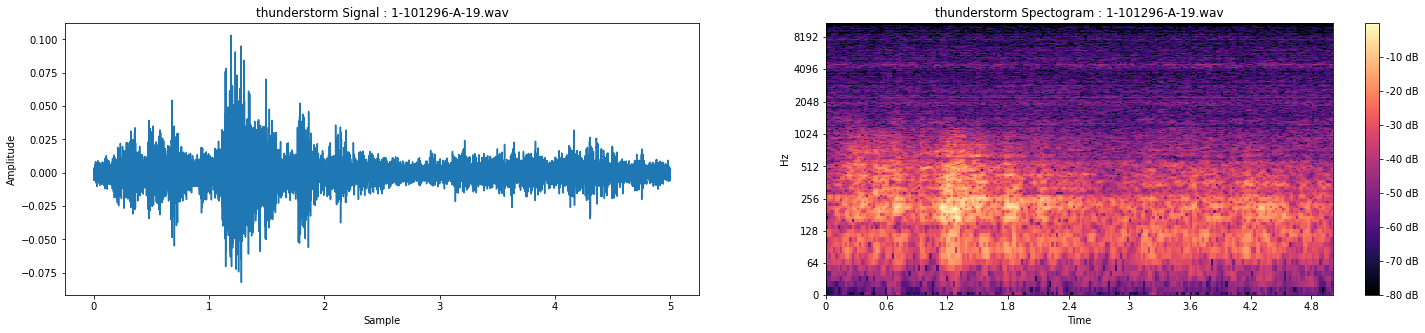}
        
        \caption{Spectrogram of an audio file of thunder}
    
    \end{subfigure}
    
    \begin{subfigure}{\textwidth}
    
        \includegraphics[width = \textwidth]{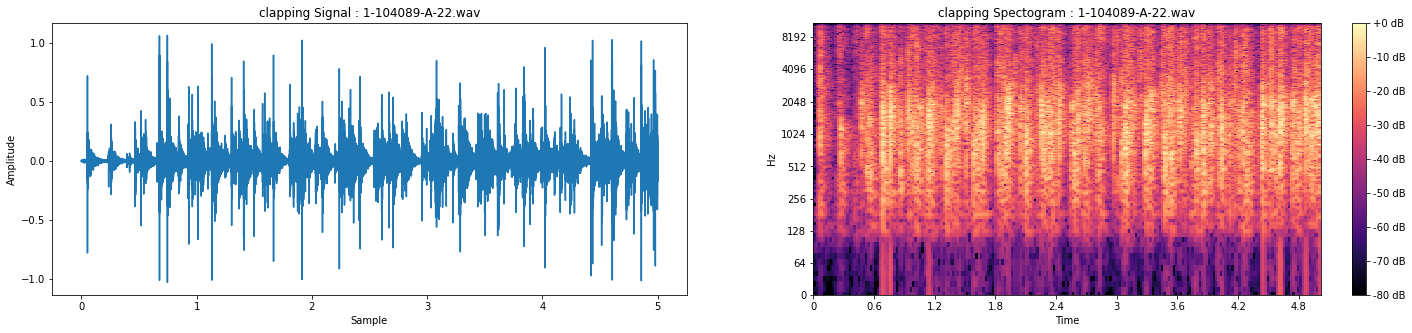}
        
        \caption{Spectrogram of an audio file of clapping}
    
    \end{subfigure}

    \begin{subfigure}{\textwidth}
    
        \includegraphics[width = \textwidth]{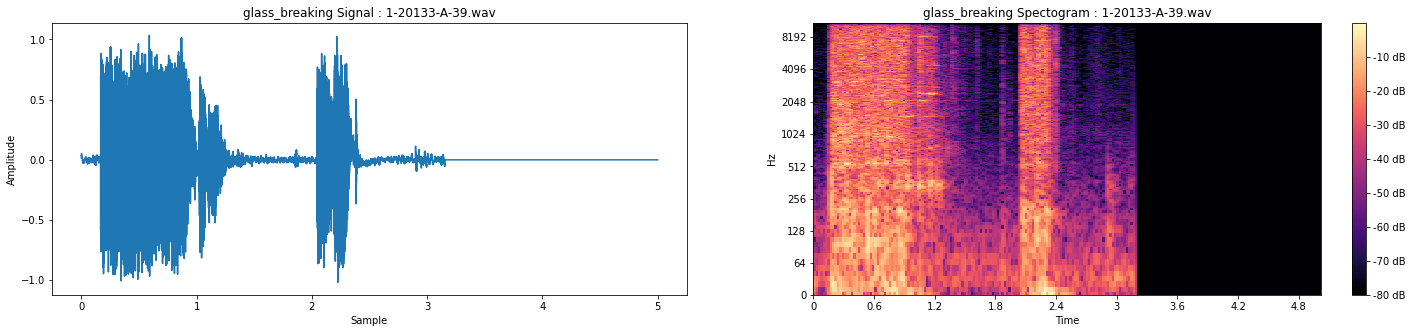}
        
        \caption{Spectrogram of an audio file of glass breaking}
    
    \end{subfigure}

    \caption{Spectrograms of some Audio files}
    \label{fig:spectrograms}
    
\end{figure}

\begin{figure}[h]
    \centering
    
    \includegraphics[width = \textwidth]{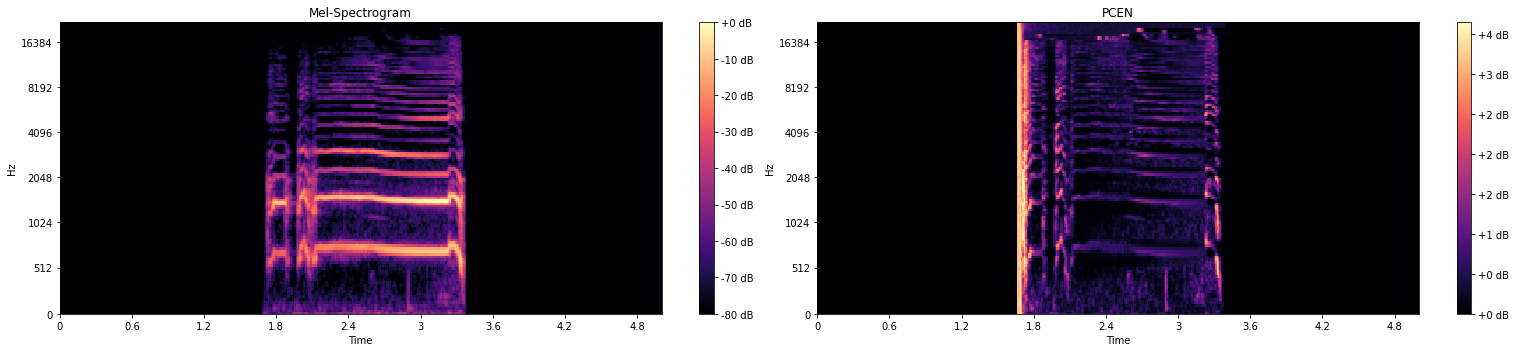}
    \includegraphics[width = \textwidth]{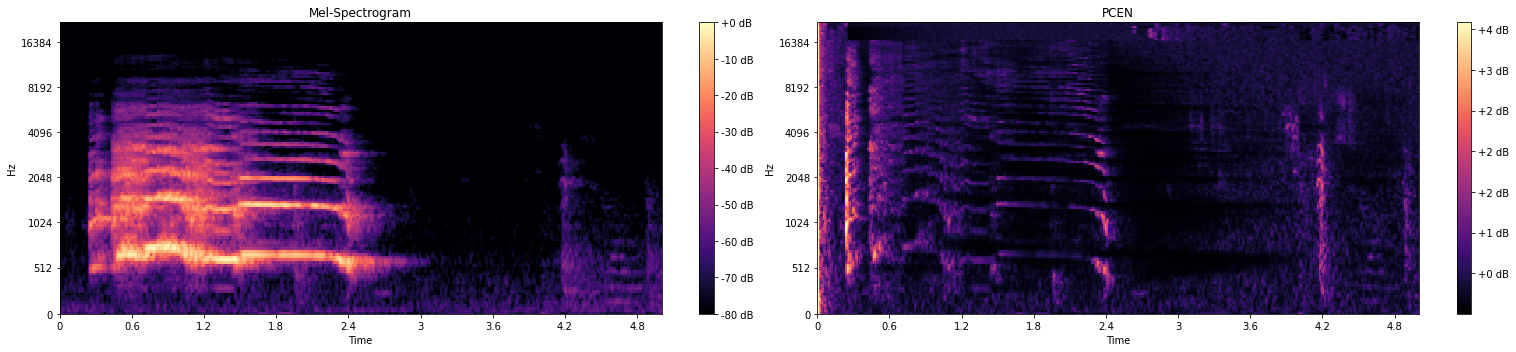}
    \caption{Spectrogram representation of PCEN done on two audio files}
    \label{fig:pcen}

\end{figure}

\begin{figure}[h]
    \centering
    
    \includegraphics[width = \textwidth]{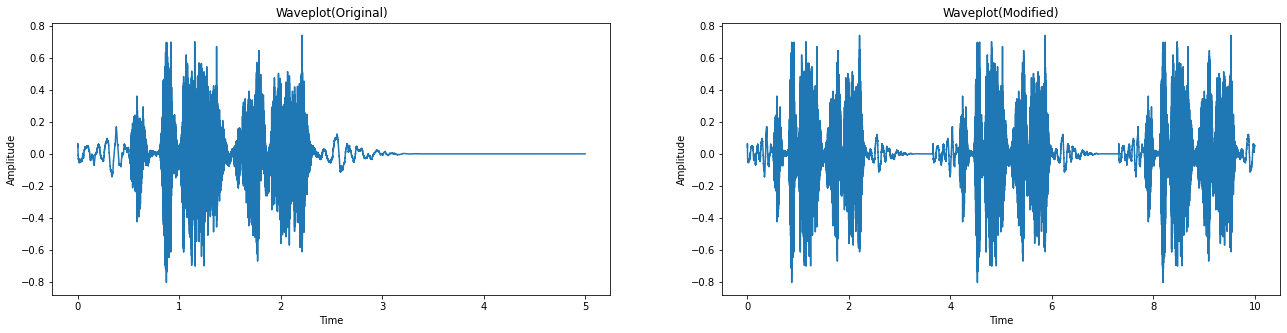}
    
    \includegraphics[width = \textwidth]{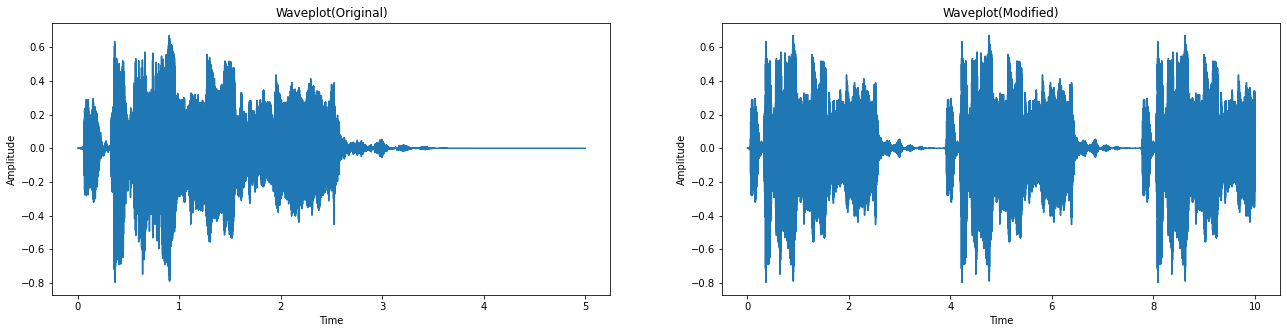}
    
    \caption{Audio Cropping done on two audio files}
    
    \label{fig:aud_crop}
    
\end{figure}

These raw spectrograms shown in Fig \ref{fig:spectrograms} are given as input to the CNN models to examine their performances on the audio files, which are shown in the column "No Filter" in the tables showing the results of the classifications performed. The spectrograms obtained with the different audio modifiers like Spectral Gating, PCEN and Audio Crop are shown in Fig \ref{fig:sg}, \ref{fig:pcen} and \ref{fig:aud_crop}, respectively. The CNN models when combined with these audio filters, use these generated spectrograms as input and the results are shown in the columns "Spectral Gating", "PCEN" and "Audio Crop" in the tables showing the results of the classifications performed. 

Based on the observations obtained from the raw spectrograms, we have fixed the lower threshold and higher threshold frequencies to 512 and 2048 Hz, respectively. The obtained spectrograms after using Low Pass Filter (threshold = 512 Hz), High Pass Filter (threshold = 2048 Hz), Band Pass Filter (lower threshold = 512 Hz and higher threshold = 2048 Hz) and Band Stop Filter (lower threshold = 512 Hz and higher threshold = 2048 Hz) are passed as input to the CNN models and the obtained results are shown in the columns Low Pass Filter, High Pass Filter, Band Pass Filter and Band Stop Filter in the tables showing the results of the classifications performed. 

\subsection{Level 1 Classification}

Now, from the results of the Level 1 Classification as shown in table \ref{tab:level-1-classifier} it can be seen that, Classification Accuracy was 75.94\% from the CNN model, EfficientNetB1 without any filtration; it increased to 74.75\% from CNN model, EfficientNetB0 with Noise Removal, then it decreased to 50.50\% with the application of PCEN; it further increased to 78.75\% with the combination of EfficientNetB2 and Audio Crop, which is the maximum  Classification Accuracy obtained in the Level 1 Classification. But then again it decreased to 48.50\% with the application of Low Pass Filter, increased to 60.25\% with High Pass Filter. Before giving the final accuracy as 36.50\% with Band Stop Filter, it showed 55.25\% in the case of Band Pass Filter.


\begin{table}[h]
\centering
\caption{Level-1 Classifier}
\label{tab:level-1-classifier}
\resizebox{\textwidth}{!}{%
\begin{tabular}{cc|cccccccc|}
\hline
\multicolumn{2}{|c|}{\textbf{Filtration Mode}} &
  \multicolumn{1}{c|}{\textbf{No Filter}} &
  \multicolumn{1}{c|}{\textbf{Noise Removal}} &
  \multicolumn{1}{c|}{\textbf{PCEN}} &
  \multicolumn{1}{c|}{\textbf{Audio Crop}} &
  \multicolumn{1}{c|}{\textbf{Low Pass Filter}} &
  \multicolumn{1}{c|}{\textbf{High Pass Filter}} &
  \multicolumn{1}{c|}{\textbf{Band Pass Filter}} &
  \multicolumn{1}{c|}{\textbf{Band Stop Filter}} \\ \hline
\multicolumn{1}{|c|}{\textbf{CNN Models}} &
  \textbf{Accuracies} &
   &
   &
   &
   &
   &
   &
   &
   \\ \cline{1-2}
\multicolumn{1}{|c|}{\multirow{2}{*}{\textbf{VGG16}}} &
  \textbf{Highest Validation Accuracy} &
  68.75\% &
  62.81\% &
  45.63\% &
  70.63\% &
  45.00\% &
  49.38\% &
  49.38\% &
  33.75\% \\ \cline{2-2}
\multicolumn{1}{|c|}{} &
  \textbf{Classification Accuracy} &
  71.00\% &
  66.00\% &
  45.00\% &
  71.50\% &
  43.75\% &
  50.50\% &
  49.50\% &
  31.25\% \\ \cline{1-2}
\multicolumn{1}{|c|}{\multirow{2}{*}{\textbf{VGG19}}} &
  \textbf{Highest Validation Accuracy} &
  67.50\% &
  68.12\% &
  43.13\% &
  72.50\% &
  48.75\% &
  47.50\% &
  52.81\% &
  35.94\% \\ \cline{2-2}
\multicolumn{1}{|c|}{} &
  \textbf{Classification Accuracy} &
  70.50\% &
  70.50\% &
  44.75\% &
  70.50\% &
  42.00\% &
  49.50\% &
  50.75\% &
  34.00\% \\ \cline{1-2}
\multicolumn{1}{|c|}{\multirow{2}{*}{\textbf{ResNet50}}} &
  \textbf{Highest Validation Accuracy} &
  71.88\% &
  64.38\% &
  43.13\% &
  80.00\% &
  49.69\% &
  49.06\% &
  47.19\% &
  37.50\% \\ \cline{2-2}
\multicolumn{1}{|c|}{} &
  \textbf{Classification Accuracy} &
  71.25\% &
  64.50\% &
  47.00\% &
  79.00\% &
  \textbf{48.50\%} &
  56.25\% &
  46.50\% &
  33.75\% \\ \cline{1-2}
\multicolumn{1}{|c|}{\multirow{2}{*}{\textbf{ResNet101}}} &
  \textbf{Highest Validation Accuracy} &
  70.31\% &
  67.19\% &
  45.31\% &
  73.75\% &
  48.44\% &
  43.75\% &
  48.75\% &
  36.56\% \\ \cline{2-2}
\multicolumn{1}{|c|}{} &
  \textbf{Classification Accuracy} &
  73.50\% &
  64.50\% &
  47.75\% &
  70.00\% &
  46.00\% &
  45.25\% &
  52.00\% &
  34.75\% \\ \cline{1-2}
\multicolumn{1}{|c|}{\multirow{2}{*}{\textbf{ResNet152}}} &
  \textbf{Highest Validation Accuracy} &
  67.50\% &
  67.19\% &
  51.88\% &
  75.00\% &
  49.38\% &
  48.75\% &
  49.69\% &
  37.81\% \\ \cline{2-2}
\multicolumn{1}{|c|}{} &
  \textbf{Classification Accuracy} &
  69.50\% &
  68.75\% &
  \textbf{50.50\%} &
  71.25\% &
  45.25\% &
  51.25\% &
  52.50\% &
  \textbf{36.50\%} \\ \cline{1-2}
\multicolumn{1}{|c|}{\multirow{2}{*}{\textbf{EfficientNetB0}}} &
  \textbf{Highest Validation Accuracy} &
  71.25\% &
  75.31\% &
  44.06\% &
  78.44\% &
  49.69\% &
  53.75\% &
  51.56\% &
  38.75\% \\ \cline{2-2}
\multicolumn{1}{|c|}{} &
  \textbf{Classification Accuracy} &
  75.50\% &
  \textbf{74.75\%} &
  47.25\% &
  76.00\% &
  47.25\% &
  56.75\% &
  53.00\% &
  31.00\% \\ \cline{1-2}
\multicolumn{1}{|c|}{\multirow{2}{*}{\textbf{EfficientNetB1}}} &
  \textbf{Highest Validation Accuracy} &
  \textbf{75.94\%} &
  70.63\% &
  41.25\% &
  75.94\% &
  48.75\% &
  55.62\% &
  56.25\% &
  38.44\% \\ \cline{2-2}
\multicolumn{1}{|c|}{} &
  \textbf{Classification Accuracy} &
  76.75\% &
  73.25\% &
  43.25\% &
  75.50\% &
  47.50\% &
  56.25\% &
  \textbf{55.25\%} &
  32.50\% \\ \cline{1-2}
\multicolumn{1}{|c|}{\multirow{2}{*}{\textbf{EfficientNetB2}}} &
  \textbf{Highest Validation Accuracy} &
  72.19\% &
  70.31\% &
  46.56\% &
  79.06\% &
  46.88\% &
  56.25\% &
  54.69\% &
  36.88\% \\ \cline{2-2}
\multicolumn{1}{|c|}{} &
  \textbf{Classification Accuracy} &
  76.25\% &
  72.25\% &
  45.25\% &
  \textbf{78.75\%} &
  45.50\% &
  \textbf{60.25\%} &
  53.00\% &
  32.25\% \\ \cline{1-2}
\multicolumn{1}{|c|}{\multirow{2}{*}{\textbf{EfficientNetB3}}} &
  \textbf{Highest Validation Accuracy} &
  72.50\% &
  69.69\% &
  43.75\% &
  74.69\% &
  53.75\% &
  55.31\% &
  56.88\% &
  36.25\% \\ \cline{2-2}
\multicolumn{1}{|c|}{} &
  \textbf{Classification Accuracy} &
  76.00\% &
  74.25\% &
  47.00\% &
  77.75\% &
  46.50\% &
  58.00\% &
  53.75\% &
  36.00\% \\ \cline{1-2}
\multicolumn{1}{|c|}{\multirow{2}{*}{\textbf{EfficientNetB4}}} &
  \textbf{Highest Validation Accuracy} &
  73.44\% &
  71.25\% &
  40.00\% &
  75.00\% &
  47.19\% &
  53.44\% &
  53.75\% &
  39.06\% \\ \cline{2-2}
\multicolumn{1}{|c|}{} &
  \textbf{Classification Accuracy} &
  77.25\% &
  70.75\% &
  41.00\% &
  74.00\% &
  47.00\% &
  55.50\% &
  52.50\% &
  32.50\% \\
  \hline
\end{tabular}%
}
\vspace{.1mm}
\end{table}

\subsection{Level 2 Classification}

\subsubsection{Animal}
\label{subsubsec:animal}

In case of Level 2 Classification of the Animal class as shown in table \ref{tab:animal-results}, the validation score started from 86.27\% from ResNet50 and ResNet152 with the raw spectrograms of the unfiltered audio files. The validation score then started to decrease to 64.71\% with PCEN, after giving the accuarcy as 82.35\% with the combination of Noise Removal and EfficientNetB3. But, again increased to 88.24\% with Audio Crop and EfficientNetB2.

After this the classifier did not increase any more and gave the accuracy results as 66.67\%, 74.51\% and 49.02\% with the audio filters. Hence, the Level 2 Classifier of Animal also got the highest accuracy from Audio crop and EfficientNetB2 like the Level 1 Classifier.

\subsubsection{Bird}
\label{subsubsec:bird}

The results of Level 2 Classification of the Bird class from the table \ref{tab:bird-results} show that here the highest validation accuracy was obtained as 96.00\% with the following combinations of audio modifier \& CNN model.

\begin{enumerate}
    \item No Filter \& VGG16
    \item No Filter \& EfficientNetB2
    \item Noise Removal \& EfficientNetB1
    \item Audio Crop \& ResNet152
    \item Audio Crop \& EfficientNetB1
    \item High Pass Filter \& ResNet50
\end{enumerate}

The accuracy was also obtained as 80.00\% with Low Pass Filter and Band Pass Filter. The least accuracy score was obtained as 64.00\% with Band Stop Filter.

\begin{table}[H]
\centering
\caption{Results for Animal class}
\label{tab:animal-results}
\resizebox{\textwidth}{!}{%
\begin{tabular}{|cc|cccccccc|}
\hline
\multicolumn{2}{|c|}{\textbf{Filtration Mode}} &
  \multicolumn{1}{c|}{\multirow{2}{*}{\textbf{No Filter}}} &
  \multicolumn{1}{c|}{\multirow{2}{*}{\textbf{Noise Removal}}} &
  \multicolumn{1}{c|}{\multirow{2}{*}{\textbf{PCEN}}} &
  \multicolumn{1}{c|}{\multirow{2}{*}{\textbf{Audio Crop}}} &
  \multicolumn{1}{c|}{\multirow{2}{*}{\textbf{Low Pass Filter}}} &
  \multicolumn{1}{c|}{\multirow{2}{*}{\textbf{High Pass Filter}}} &
  \multicolumn{1}{c|}{\multirow{2}{*}{\textbf{Band Pass Filter}}} &
  \multicolumn{1}{c|}{\multirow{2}{*}{\textbf{Band Stop Filter}}} \\ \cline{1-2}
\multicolumn{1}{|c|}{\textbf{CNN Models}} &
  \textbf{Accuracies} &
  \multicolumn{1}{c|}{} &
  \multicolumn{1}{c|}{} &
  \multicolumn{1}{c|}{} &
  \multicolumn{1}{c|}{} &
  \multicolumn{1}{c|}{} &
  \multicolumn{1}{c|}{} &
  \multicolumn{1}{c|}{} &
  \multicolumn{1}{c|}{} \\ \hline
\multicolumn{1}{|c|}{\multirow{2}{*}{\textbf{VGG16}}} &
  \textbf{Highest Validation Accuracy} &
  66.67\% &
  70.59\% &
  58.82\% &
  76.47\% &
  64.71\% &
  50.98\% &
  49.02\% &
  47.06\% \\ \cline{2-2}
\multicolumn{1}{|c|}{} &
  \textbf{Classification Accuracy} &
  79.69\% &
  71.88\% &
  50.00\% &
  70.31\% &
  48.44\% &
  48.44\% &
  54.69\% &
  42.19\% \\ \cline{1-2}
\multicolumn{1}{|c|}{\multirow{2}{*}{\textbf{VGG19}}} &
  \textbf{Highest Validation Accuracy} &
  76.47\% &
  68.63\% &
  43.14\% &
  76.47\% &
  58.82\% &
  64.71\% &
  70.59\% &
  39.22\% \\ \cline{2-2}
\multicolumn{1}{|c|}{} &
  \textbf{Classification Accuracy} &
  81.25\% &
  71.88\% &
  39.06\% &
  73.44\% &
  46.88\% &
  67.19\% &
  48.44\% &
  35.94\% \\ \cline{1-2}
\multicolumn{1}{|c|}{\multirow{2}{*}{\textbf{ResNet50}}} &
  \textbf{Highest Validation Accuracy} &
  \textbf{86.27\%} &
  76.47\% &
  49.02\% &
  84.31\% &
  60.78\% &
  60.78\% &
  72.55\% &
  35.29\% \\ \cline{2-2}
\multicolumn{1}{|c|}{} &
  \textbf{Classification Accuracy} &
  82.81\% &
  84.38\% &
  46.88\% &
  79.69\% &
  34.38\% &
  48.44\% &
  68.75\% &
  29.69\% \\ \cline{1-2}
\multicolumn{1}{|c|}{\multirow{2}{*}{\textbf{ResNet101}}} &
  \textbf{Highest Validation Accuracy} &
  76.47\% &
  74.51\% &
  50.98\% &
  82.35\% &
  \textbf{66.67\%} &
  64.71\% &
  58.82\% &
  35.29\% \\ \cline{2-2}
\multicolumn{1}{|c|}{} &
  \textbf{Classification Accuracy} &
  75.00\% &
  73.44\% &
  34.38\% &
  81.25\% &
  56.25\% &
  57.81\% &
  62.50\% &
  23.44\% \\ \cline{1-2}
\multicolumn{1}{|c|}{\multirow{2}{*}{\textbf{ResNet152}}} &
  \textbf{Highest Validation Accuracy} &
  \textbf{86.27\%} &
  76.47\% &
  56.86\% &
  86.27\% &
  62.75\% &
  \textbf{74.51\%} &
  66.67\% &
  43.14\% \\ \cline{2-2}
\multicolumn{1}{|c|}{} &
  \textbf{Classification Accuracy} &
  84.38\% &
  75.00\% &
  53.12\% &
  78.12\% &
  50.00\% &
  67.19\% &
  71.88\% &
  34.38\% \\ \cline{1-2}
\multicolumn{1}{|c|}{\multirow{2}{*}{\textbf{EfficientNetB0}}} &
  \textbf{Highest Validation Accuracy} &
  82.35\% &
  72.55\% &
  \textbf{64.71\%} &
  76.47\% &
  62.75\% &
  64.71\% &
  62.75\% &
  39.22\% \\ \cline{2-2}
\multicolumn{1}{|c|}{} &
  \textbf{Classification Accuracy} &
  84.38\% &
  75.00\% &
  54.69\% &
  78.12\% &
  57.81\% &
  70.31\% &
  54.69\% &
  35.94\% \\ \cline{1-2}
\multicolumn{1}{|c|}{\multirow{2}{*}{\textbf{EfficientNetB1}}} &
  \textbf{Highest Validation Accuracy} &
  84.31\% &
  70.59\% &
  56.86\% &
  82.35\% &
  60.78\% &
  70.59\% &
  \textbf{74.51\%} &
  45.10\% \\ \cline{2-2}
\multicolumn{1}{|c|}{} &
  \textbf{Classification Accuracy} &
  78.12\% &
  79.69\% &
  50.00\% &
  82.81\% &
  62.50\% &
  70.31\% &
  67.19\% &
  39.06\% \\ \cline{1-2}
\multicolumn{1}{|c|}{\multirow{2}{*}{\textbf{EfficientNetB2}}} &
  \textbf{Highest Validation Accuracy} &
  80.39\% &
  66.67\% &
  58.82\% &
  \textbf{88.24\%} &
  \textbf{66.67\%} &
  \textbf{74.51\%} &
  66.67\% &
  33.33\% \\ \cline{2-2}
\multicolumn{1}{|c|}{} &
  \textbf{Classification Accuracy} &
  79.69\% &
  76.56\% &
  65.62\% &
  82.81\% &
  59.38\% &
  64.06\% &
  62.50\% &
  37.50\% \\ \cline{1-2}
\multicolumn{1}{|c|}{\multirow{2}{*}{\textbf{EfficientNetB3}}} &
  \textbf{Highest Validation Accuracy} &
  84.31\% &
  \textbf{82.35\%} &
  60.78\% &
  70.59\% &
  \textbf{66.67\%} &
  58.82\% &
  70.59\% &
  \textbf{49.02\%} \\ \cline{2-2}
\multicolumn{1}{|c|}{} &
  \textbf{Classification Accuracy} &
  89.06\% &
  81.25\% &
  54.69\% &
  81.25\% &
  60.94\% &
  65.62\% &
  64.06\% &
  34.38\% \\ \cline{1-2}
\multicolumn{1}{|c|}{\multirow{2}{*}{\textbf{EfficientNetB4}}} &
  \textbf{Highest Validation Accuracy} &
  82.35\% &
  68.63\% &
  60.78\% &
  76.47\% &
  62.75\% &
  60.78\% &
  66.67\% &
  31.37\% \\ \cline{2-2}
\multicolumn{1}{|c|}{} &
  \textbf{Classification Accuracy} &
  79.69\% &
  73.44\% &
  53.12\% &
  81.25\% &
  51.56\% &
  68.75\% &
  62.50\% &
  29.69\% \\ 
  \hline
\end{tabular}%
}
\end{table}

\begin{table}[H]
\centering
\caption{Results for Bird class}
\label{tab:bird-results}
\resizebox{\textwidth}{!}{%
\begin{tabular}{|cc|cccccccc|}
\hline
\multicolumn{2}{|c|}{\textbf{Filtration Mode}} &
  \multicolumn{1}{c|}{\multirow{2}{*}{\textbf{No Filter}}} &
  \multicolumn{1}{c|}{\multirow{2}{*}{\textbf{Noise Removal}}} &
  \multicolumn{1}{c|}{\multirow{2}{*}{\textbf{PCEN}}} &
  \multicolumn{1}{c|}{\multirow{2}{*}{\textbf{Audio Crop}}} &
  \multicolumn{1}{c|}{\multirow{2}{*}{\textbf{Low Pass Filter}}} &
  \multicolumn{1}{c|}{\multirow{2}{*}{\textbf{High Pass Filter}}} &
  \multicolumn{1}{c|}{\multirow{2}{*}{\textbf{Band Pass Filter}}} &
  \multicolumn{1}{c|}{\multirow{2}{*}{\textbf{Band Stop Filter}}} \\ \cline{1-2}
\multicolumn{1}{|c|}{\textbf{CNN Models}} &
  \textbf{Accuracies} &
  \multicolumn{1}{c|}{} &
  \multicolumn{1}{c|}{} &
  \multicolumn{1}{c|}{} &
  \multicolumn{1}{c|}{} &
  \multicolumn{1}{c|}{} &
  \multicolumn{1}{c|}{} &
  \multicolumn{1}{c|}{} &
  \multicolumn{1}{c|}{} \\ \hline
\multicolumn{1}{|c|}{\multirow{2}{*}{\textbf{VGG16}}} &
  \textbf{Highest Validation Accuracy} &
  \textbf{96.00\%} &
  88.00\% &
  40.00\% &
  84.00\% &
  72.00\% &
  88.00\% &
  76.00\% &
  64.00\% \\ \cline{2-2}
\multicolumn{1}{|c|}{} &
  \textbf{Classification Accuracy} &
  87.50\% &
  84.38\% &
  59.38\% &
  87.50\% &
  59.38\% &
  75.00\% &
  68.75\% &
  59.38\% \\ \cline{1-2}
\multicolumn{1}{|c|}{\multirow{2}{*}{\textbf{VGG19}}} &
  \textbf{Highest Validation Accuracy} &
  88.00\% &
  92.00\% &
  68.00\% &
  92.00\% &
  68.00\% &
  84.00\% &
  76.00\% &
  60.00\% \\ \cline{2-2}
\multicolumn{1}{|c|}{} &
  \textbf{Classification Accuracy} &
  75.00\% &
  78.12\% &
  68.75\% &
  81.25\% &
  56.25\% &
  75.00\% &
  81.25\% &
  68.75\% \\ \cline{1-2}
\multicolumn{1}{|c|}{\multirow{2}{*}{\textbf{ResNet50}}} &
  \textbf{Highest Validation Accuracy} &
  92.00\% &
  92.00\% &
  48.00\% &
  80.00\% &
  \textbf{80.00\%} &
  \textbf{96.00\%} &
  76.00\% &
  64.00\% \\ \cline{2-2}
\multicolumn{1}{|c|}{} &
  \textbf{Classification Accuracy} &
  84.38\% &
  93.75\% &
  78.12\% &
  78.12\% &
  68.75\% &
  81.25\% &
  68.75\% &
  56.25\% \\ \cline{1-2}
\multicolumn{1}{|c|}{\multirow{2}{*}{\textbf{ResNet101}}} &
  \textbf{Highest Validation Accuracy} &
  88.00\% &
  92.00\% &
  56.00\% &
  88.00\% &
  72.00\% &
  84.00\% &
  72.00\% &
  60.00\% \\ \cline{2-2}
\multicolumn{1}{|c|}{} &
  \textbf{Classification Accuracy} &
  87.50\% &
  81.25\% &
  56.25\% &
  84.38\% &
  65.62\% &
  71.88\% &
  71.88\% &
  50.00\% \\ \cline{1-2}
\multicolumn{1}{|c|}{\multirow{2}{*}{\textbf{ResNet152}}} &
  \textbf{Highest Validation Accuracy} &
  92.00\% &
  88.00\% &
  44.00\% &
  \textbf{96.00\%} &
  72.00\% &
  88.00\% &
  72.00\% &
  48.00\% \\ \cline{2-2}
\multicolumn{1}{|c|}{} &
  \textbf{Classification Accuracy} &
  93.75\% &
  65.62\% &
  62.50\% &
  90.62\% &
  50.00\% &
  71.88\% &
  75.00\% &
  50.00\% \\ \cline{1-2}
\multicolumn{1}{|c|}{\multirow{2}{*}{\textbf{EfficientNetB0}}} &
  \textbf{Highest Validation Accuracy} &
  92.00\% &
  84.00\% &
  64.00\% &
  92.00\% &
  72.00\% &
  88.00\% &
  \textbf{80.00\%} &
  60.00\% \\ \cline{2-2}
\multicolumn{1}{|c|}{} &
  \textbf{Classification Accuracy} &
  92.00\% &
  78.12\% &
  71.88\% &
  81.25\% &
  56.25\% &
  87.50\% &
  68.75\% &
  56.25\% \\ \cline{1-2}
\multicolumn{1}{|c|}{\multirow{2}{*}{\textbf{EfficientNetB1}}} &
  \textbf{Highest Validation Accuracy} &
  92.00\% &
  \textbf{96.00\%} &
  68.00\% &
  \textbf{96.00\%} &
  68.00\% &
  92.00\% &
  72.00\% &
  60.00\% \\ \cline{2-2}
\multicolumn{1}{|c|}{} &
  \textbf{Classification Accuracy} &
  78.12\% &
  81.25\% &
  68.75\% &
  87.50\% &
  65.62\% &
  62.50\% &
  78.12\% &
  65.62\% \\ \cline{1-2}
\multicolumn{1}{|c|}{\multirow{2}{*}{\textbf{EfficientNetB2}}} &
  \textbf{Highest Validation Accuracy} &
  \textbf{96.00\%} &
  80.00\% &
  52.00\% &
  92.00\% &
  68.00\% &
  80.00\% &
  68.00\% &
  \textbf{64.00\%} \\ \cline{2-2}
\multicolumn{1}{|c|}{} &
  \textbf{Classification Accuracy} &
  90.62\% &
  81.25\% &
  68.75\% &
  84.38\% &
  56.25\% &
  78.12\% &
  71.88\% &
  62.50\% \\ \cline{1-2}
\multicolumn{1}{|c|}{\multirow{2}{*}{\textbf{EfficientNetB3}}} &
  \textbf{Highest Validation Accuracy} &
  88.00\% &
  84.00\% &
  \textbf{72.00\%} &
  88.00\% &
  68.00\% &
  92.00\% &
  76.00\% &
  60.00\% \\ \cline{2-2}
\multicolumn{1}{|c|}{} &
  \textbf{Classification Accuracy} &
  81.25\% &
  78.12\% &
  65.62\% &
  81.25\% &
  65.62\% &
  78.12\% &
  68.75\% &
  65.62\% \\ \cline{1-2}
\multicolumn{1}{|c|}{\multirow{2}{*}{\textbf{EfficientNetB4}}} &
  \textbf{Highest Validation Accuracy} &
  92.00\% &
  84.00\% &
  68.00\% &
  88.00\% &
  76.00\% &
  84.00\% &
  76.00\% &
  64.00\% \\ \cline{2-2}
\multicolumn{1}{|c|}{} &
  \textbf{Classification Accuracy} &
  87.50\% &
  78.12\% &
  71.88\% &
  81.25\% &
  62.50\% &
  71.88\% &
  81.25\% &
  62.50\% \\ 
  \hline
\end{tabular}%
}
\end{table}

\subsubsection{Natural Soundscapes}
\label{subsubsec:natural-soundscapes}

Finally, from the results table \ref{tab:nature-results} of the Level 2 Classifier of Nature class it is clear that, the highest validation accuracy obtained is 95.45\% from the application combinations of No Filter \& ResNet152, Noise Removal \& ResNet152 and Noise Removal \& EfficientNetB1. Here, the accuracy score was also obtained as high as 90.91\% with Band Pass Filter and also 86.36\% with the applications of Audio Crop and PCEN. But, in this case the minimum highest validation accuracy was obtained from the application of Low Pass Filter as 54.55\%.

\begin{table}[H]
\centering
\caption{Results for Nature class}
\label{tab:nature-results}
\resizebox{\textwidth}{!}{%
\begin{tabular}{|cc|cccccccc|}
\hline
\multicolumn{2}{|c|}{\textbf{Filtration Mode}} &
  \multicolumn{1}{c|}{\multirow{2}{*}{\textbf{No Filter}}} &
  \multicolumn{1}{c|}{\multirow{2}{*}{\textbf{Noise Removal}}} &
  \multicolumn{1}{c|}{\multirow{2}{*}{\textbf{PCEN}}} &
  \multicolumn{1}{c|}{\multirow{2}{*}{\textbf{Audio Crop}}} &
  \multicolumn{1}{c|}{\multirow{2}{*}{\textbf{Low Pass Filter}}} &
  \multicolumn{1}{c|}{\multirow{2}{*}{\textbf{High Pass Filter}}} &
  \multicolumn{1}{c|}{\multirow{2}{*}{\textbf{Band Pass Filter}}} &
  \multicolumn{1}{c|}{\multirow{2}{*}{\textbf{Band Stop Filter}}} \\ \cline{1-2}
\multicolumn{1}{|c|}{\textbf{CNN Models}} &
  \textbf{Accuracies} &
  \multicolumn{1}{c|}{} &
  \multicolumn{1}{c|}{} &
  \multicolumn{1}{c|}{} &
  \multicolumn{1}{c|}{} &
  \multicolumn{1}{c|}{} &
  \multicolumn{1}{c|}{} &
  \multicolumn{1}{c|}{} &
  \multicolumn{1}{c|}{} \\ \hline
\multicolumn{1}{|c|}{\multirow{2}{*}{\textbf{VGG16}}} &
  \textbf{Highest Validation Accuracy} &
  84.09\% &
  86.36\% &
  75.00\% &
  \textbf{86.36\%} &
  47.73\% &
  75.00\% &
  77.27\% &
  45.45\% \\ \cline{2-2}
\multicolumn{1}{|c|}{} &
  \textbf{Classification Accuracy} &
  75.00\% &
  85.71\% &
  60.71\% &
  83.93\% &
  69.64\% &
  82.14\% &
  69.64\% &
  41.07\% \\ \cline{1-2}
\multicolumn{1}{|c|}{\multirow{2}{*}{\textbf{VGG19}}} &
  \textbf{Highest Validation Accuracy} &
  86.36\% &
  93.18\% &
  63.64\% &
  84.09\% &
  52.27\% &
  77.27\% &
  75.00\% &
  45.45\% \\ \cline{2-2}
\multicolumn{1}{|c|}{} &
  \textbf{Classification Accuracy} &
  82.14\% &
  85.71\% &
  60.71\% &
  87.50\% &
  53.57\% &
  69.64\% &
  69.64\% &
  39.29\% \\ \cline{1-2}
\multicolumn{1}{|c|}{\multirow{2}{*}{\textbf{ResNet50}}} &
  \textbf{Highest Validation Accuracy} &
  90.91\% &
  93.18\% &
  70.45\% &
  77.27\% &
  52.27\% &
  79.55\% &
  84.09\% &
  47.73\% \\ \cline{2-2}
\multicolumn{1}{|c|}{} &
  \textbf{Classification Accuracy} &
  83.93\% &
  92.86\% &
  50.00\% &
  87.50\% &
  73.21\% &
  71.43\% &
  66.07\% &
  39.29\% \\ \cline{1-2}
\multicolumn{1}{|c|}{\multirow{2}{*}{\textbf{ResNet101}}} &
  \textbf{Highest Validation Accuracy} &
  88.64\% &
  88.64\% &
  77.27\% &
  \textbf{86.36\%} &
  50.00\% &
  \textbf{84.09\%} &
  72.73\% &
  47.73\% \\ \cline{2-2}
\multicolumn{1}{|c|}{} &
  \textbf{Classification Accuracy} &
  83.93\% &
  87.50\% &
  60.71\% &
  85.71\% &
  64.29\% &
  75.00\% &
  66.07\% &
  26.79\% \\ \cline{1-2}
\multicolumn{1}{|c|}{\multirow{2}{*}{\textbf{ResNet152}}} &
  \textbf{Highest Validation Accuracy} &
  \textbf{95.45\%} &
  \textbf{95.45\%} &
  81.82\% &
  \textbf{86.36\%} &
  52.27\% &
  77.27\% &
  \textbf{90.91\%} &
  47.73\% \\ \cline{2-2}
\multicolumn{1}{|c|}{} &
  \textbf{Classification Accuracy} &
  85.71\% &
  92.86\% &
  60.71\% &
  87.50\% &
  71.43\% &
  73.21\% &
  67.86\% &
  26.79\% \\ \cline{1-2}
\multicolumn{1}{|c|}{\multirow{2}{*}{\textbf{EfficientNetB0}}} &
  \textbf{Highest Validation Accuracy} &
  88.64\% &
  90.91\% &
  81.82\% &
  84.09\% &
  \textbf{54.55\%} &
  77.27\% &
  81.82\% &
  \textbf{56.82\%} \\ \cline{2-2}
\multicolumn{1}{|c|}{} &
  \textbf{Classification Accuracy} &
  83.93\% &
  94.64\% &
  67.86\% &
  89.29\% &
  66.07\% &
  76.79\% &
  67.86\% &
  46.43\% \\ \cline{1-2}
\multicolumn{1}{|c|}{\multirow{2}{*}{\textbf{EfficientNetB1}}} &
  \textbf{Highest Validation Accuracy} &
  90.91\% &
  \textbf{95.45\%} &
  81.82\% &
  84.09\% &
  45.45\% &
  \textbf{84.09\%} &
  77.27\% &
  47.73\% \\ \cline{2-2}
\multicolumn{1}{|c|}{} &
  \textbf{Classification Accuracy} &
  85.71\% &
  94.64\% &
  66.07\% &
  85.71\% &
  69.64\% &
  80.36\% &
  71.43\% &
  48.21\% \\ \cline{1-2}
\multicolumn{1}{|c|}{\multirow{2}{*}{\textbf{EfficientNetB2}}} &
  \textbf{Highest Validation Accuracy} &
  88.64\% &
  90.91\% &
  84.09\% &
  81.82\% &
  43.18\% &
  65.91\% &
  79.55\% &
  52.27\% \\ \cline{2-2}
\multicolumn{1}{|c|}{} &
  \textbf{Classification Accuracy} &
  78.57\% &
  92.86\% &
  73.21\% &
  85.71\% &
  66.07\% &
  67.86\% &
  73.21\% &
  51.79\% \\ \cline{1-2}
\multicolumn{1}{|c|}{\multirow{2}{*}{\textbf{EfficientNetB3}}} &
  \textbf{Highest Validation Accuracy} &
  84.09\% &
  90.91\% &
  \textbf{86.36\%} &
  84.09\% &
  47.73\% &
  86.36\% &
  79.55\% &
  54.55\% \\ \cline{2-2}
\multicolumn{1}{|c|}{} &
  \textbf{Classification Accuracy} &
  75.00\% &
  92.86\% &
  66.07\% &
  85.71\% &
  71.43\% &
  82.14\% &
  75.00\% &
  46.43\% \\ \cline{1-2}
\multicolumn{1}{|c|}{\multirow{2}{*}{\textbf{EfficientNetB4}}} &
  \textbf{Highest Validation Accuracy} &
  81.82\% &
  90.91\% &
  81.82\% &
  79.55\% &
  \textbf{54.45\%} &
  \textbf{84.09\%} &
  88.64\% &
  50.00\% \\ \cline{2-2}
\multicolumn{1}{|c|}{} &
  \textbf{Classification Accuracy} &
  80.36\% &
  87.50\% &
  60.71\% &
  83.93\% &
  69.64\% &
  78.57\% &
  67.86\% &
  55.36\% \\ 
  \hline
\end{tabular}%
}
\end{table}

\begin{table}[H]
\centering
\caption{Results for Human class}
\label{tab:human-results}
\resizebox{\textwidth}{!}{%
\begin{tabular}{|cc|cccccccc|}
\hline
\multicolumn{2}{|c|}{\textbf{Filtration Mode}} &
  \multicolumn{1}{c|}{\multirow{2}{*}{\textbf{No Filter}}} &
  \multicolumn{1}{c|}{\multirow{2}{*}{\textbf{Noise Removal}}} &
  \multicolumn{1}{c|}{\multirow{2}{*}{\textbf{PCEN}}} &
  \multicolumn{1}{c|}{\multirow{2}{*}{\textbf{Audio Crop}}} &
  \multicolumn{1}{c|}{\multirow{2}{*}{\textbf{Low Pass Filter}}} &
  \multicolumn{1}{c|}{\multirow{2}{*}{\textbf{High Pass Filter}}} &
  \multicolumn{1}{c|}{\multirow{2}{*}{\textbf{Band Pass Filter}}} &
  \multicolumn{1}{c|}{\multirow{2}{*}{\textbf{Band Stop Filter}}} \\ \cline{1-2}
\multicolumn{1}{|c|}{\textbf{CNN Models}} &
  \textbf{Accuracies} &
  \multicolumn{1}{c|}{} &
  \multicolumn{1}{c|}{} &
  \multicolumn{1}{c|}{} &
  \multicolumn{1}{c|}{} &
  \multicolumn{1}{c|}{} &
  \multicolumn{1}{c|}{} &
  \multicolumn{1}{c|}{} &
  \multicolumn{1}{c|}{} \\ \hline
\multicolumn{1}{|c|}{\multirow{2}{*}{\textbf{VGG16}}} &
  \textbf{Highest Validation Accuracy} &
  75.00\% &
  71.88\% &
  53.12\% &
  76.56\% &
  56.25\% &
  56.25\% &
  67.19\% &
  43.75\% \\ \cline{2-2}
\multicolumn{1}{|c|}{} &
  \textbf{Classification Accuracy} &
  81.25\% &
  80.00\% &
  51.25\% &
  83.75\% &
  47.5\% &
  56.25\% &
  63.75\% &
  38.75\% \\ \cline{1-2}
\multicolumn{1}{|c|}{\multirow{2}{*}{\textbf{VGG19}}} &
  \textbf{Highest Validation Accuracy} &
  68.75\% &
  71.88\% &
  60.94\% &
  78.12\% &
  \textbf{62.5\%} &
  62.5\% &
  62.5\% &
  43.75\% \\ \cline{2-2}
\multicolumn{1}{|c|}{} &
  \textbf{Classification Accuracy} &
  83.75\% &
  83.75\% &
  58.75\% &
  76.25\% &
  53.75\% &
  52.5\% &
  58.75\% &
  38.75\% \\ \cline{1-2}
\multicolumn{1}{|c|}{\multirow{2}{*}{\textbf{ResNet50}}} &
  \textbf{Highest Validation Accuracy} &
  81.25\% &
  84.38\% &
  51.56\% &
  89.06\% &
  \textbf{62.5\%} &
  \textbf{67.19\%} &
  \textbf{78.12\%} &
  42.19\% \\ \cline{2-2}
\multicolumn{1}{|c|}{} &
  \textbf{Classification Accuracy} &
  83.75\% &
  81.25\% &
  53.75\% &
  81.25\% &
  52.5\% &
  63.75\% &
  67.1\% &
  42.5\% \\ \cline{1-2}
\multicolumn{1}{|c|}{\multirow{2}{*}{\textbf{ResNet101}}} &
  \textbf{Highest Validation Accuracy} &
  79.69\% &
  82.81\% &
  51.56\% &
  84.38\% &
  56.25\% &
  56.25\% &
  67.19\% &
  40.62\% \\ \cline{2-2}
\multicolumn{1}{|c|}{} &
  \textbf{Classification Accuracy} &
  83.75\% &
  86.25\% &
  43.75\% &
  82.5\% &
  50\% &
  55\% &
  65\% &
  43.75\% \\ \cline{1-2}
\multicolumn{1}{|c|}{\multirow{2}{*}{\textbf{ResNet152}}} &
  \textbf{Highest Validation Accuracy} &
  82.81\% &
  78.12\% &
  57.81\% &
  84.38\% &
  54.69\% &
  64.06\% &
  65.62\% &
  46.88\% \\ \cline{2-2}
\multicolumn{1}{|c|}{} &
  \textbf{Classification Accuracy} &
  86.25\% &
  80.00\% &
  50.00\% &
  82.5\% &
  57.5\% &
  61.25\% &
  63.75\% &
  46.25\% \\ \cline{1-2}
\multicolumn{1}{|c|}{\multirow{2}{*}{\textbf{EfficientNetB0}}} &
  \textbf{Highest Validation Accuracy} &
  81.25\% &
  \textbf{87.50\%} &
  68.75\% &
  \textbf{93.75\%} &
  56.25\% &
  65.62\% &
  65.62\% &
  \textbf{51.56\%} \\ \cline{2-2}
\multicolumn{1}{|c|}{} &
  \textbf{Classification Accuracy} &
  88.75\% &
  92.50\% &
  65.00\% &
  85\% &
  56.25\% &
  56.25\% &
  65\% &
  43.75\% \\ \cline{1-2}
\multicolumn{1}{|c|}{\multirow{2}{*}{\textbf{EfficientNetB1}}} &
  \textbf{Highest Validation Accuracy} &
  82.81\% &
  82.81\% &
  67.19\% &
  89.06\% &
  56.25\% &
  64.06\% &
  71.88\% &
  42.19\% \\ \cline{2-2}
\multicolumn{1}{|c|}{} &
  \textbf{Classification Accuracy} &
  87.50\% &
  85.00\% &
  63.75\% &
  77.5\% &
  55\% &
  65\% &
  66.25\% &
  46.25\% \\ \cline{1-2}
\multicolumn{1}{|c|}{\multirow{2}{*}{\textbf{EfficientNetB2}}} &
  \textbf{Highest Validation Accuracy} &
  81.25\% &
  85.94\% &
  67.19\% &
  90.62\% &
  60.94\% &
  62.5\% &
  68.75\% &
  \textbf{51.56\%} \\ \cline{2-2}
\multicolumn{1}{|c|}{} &
  \textbf{Classification Accuracy} &
  88.75\% &
  91.25\% &
  60.00\% &
  86.25\% &
  51.25\% &
  67.5\% &
  58.75\% &
  50\% \\ \cline{1-2}
\multicolumn{1}{|c|}{\multirow{2}{*}{\textbf{EfficientNetB3}}} &
  \textbf{Highest Validation Accuracy} &
  \textbf{84.38\%} &
  81.25\% &
  \textbf{70.31\%} &
  90.62\% &
  51.56\% &
  \textbf{67.19\%} &
  73.44\% &
  50\% \\ \cline{2-2}
\multicolumn{1}{|c|}{} &
  \textbf{Classification Accuracy} &
  91.25\% &
  86.25\% &
  57.50\% &
  86.25\% &
  53.75\% &
  63.75\% &
  67.5\% &
  48.75\% \\ \cline{1-2}
\multicolumn{1}{|c|}{\multirow{2}{*}{\textbf{EfficientNetB4}}} &
  \textbf{Highest Validation Accuracy} &
  \textbf{84.38\%} &
  82.81\% &
  64.06\% &
  84.38\% &
  57.81\% &
  62.5\% &
  68.75\% &
  45.3\% \\ \cline{2-2}
\multicolumn{1}{|c|}{} &
  \textbf{Classification Accuracy} &
  92.50\% &
  76.25\% &
  58.75\% &
  83.75\% &
  53.75\% &
  71.25\% &
  60\% &
  43.75\% \\ 
  \hline
\end{tabular}%
}
\end{table}

\subsubsection{Human}
\label{subsubsec:human}

93.75\% is the highest validation accuracy obtained by the Level 2 Classifier of the Human class as shown in the results of table \ref{tab:human-results}, with the combiation of Audio Crop and the CNN model EfficientNetB0. It also got validation score of 87.50\% with Noise Removal. But, like the Level 2 Classifiers of the previous classes it got minimum highest validation accuracy as 51.56\% with the application of Band Stop Filter.

\subsubsection{Machine Sounds}
\label{subsubsec:machine-sounds}

The classification results from the table \ref{tab:machine-sounds-results} of the Level 2 Classification for Machine Sounds show that, the classifier got highest validation accuracy as 92.00\% with the following combinations of audio modifiers \& CNN models as follows.

\begin{enumerate}
    \item No Filter \& VGG16
    \item No Filter \& ResNet152
    \item No Filter \& EfficientNetB3
    \item Audio Crop \& VGG19
    \item Audio Crop \& ResNet101
    \item Audio Crop \& ResNet152
    \item Audio Crop \& EfficientNetB3
\end{enumerate}
The minimum accuracy score obtained was 64.00\% with Band Stop Filter.

\begin{table}[H]
\centering
\caption{Results for Machine Sounds class}
\label{tab:machine-sounds-results}
\resizebox{\textwidth}{!}{%
\begin{tabular}{|cc|cccccccc|}
\hline
\multicolumn{2}{|c|}{\textbf{Filtration Mode}} &
  \multicolumn{1}{c|}{\multirow{2}{*}{\textbf{No Filter}}} &
  \multicolumn{1}{c|}{\multirow{2}{*}{\textbf{Noise Removal}}} &
  \multicolumn{1}{c|}{\multirow{2}{*}{\textbf{PCEN}}} &
  \multicolumn{1}{c|}{\multirow{2}{*}{\textbf{Audio Crop}}} &
  \multicolumn{1}{c|}{\multirow{2}{*}{\textbf{Low Pass Filter}}} &
  \multicolumn{1}{c|}{\multirow{2}{*}{\textbf{High Pass Filter}}} &
  \multicolumn{1}{c|}{\multirow{2}{*}{\textbf{Band Pass Filter}}} &
  \multicolumn{1}{c|}{\multirow{2}{*}{\textbf{Band Stop Filter}}} \\ \cline{1-2}
\multicolumn{1}{|c|}{\textbf{CNN Models}} &
  \textbf{Accuracies} &
  \multicolumn{1}{c|}{} &
  \multicolumn{1}{c|}{} &
  \multicolumn{1}{c|}{} &
  \multicolumn{1}{c|}{} &
  \multicolumn{1}{c|}{} &
  \multicolumn{1}{c|}{} &
  \multicolumn{1}{c|}{} &
  \multicolumn{1}{c|}{} \\ \hline
\multicolumn{1}{|c|}{\multirow{2}{*}{\textbf{VGG16}}} &
  \textbf{Highest Validation Accuracy} &
  \textbf{92.00\%} &
  \textbf{84.00\%} &
  60.00\% &
  96.00\% &
  64.00\% &
  80.00\% &
  68.00\% &
  60.00\% \\ \cline{2-2}
\multicolumn{1}{|c|}{} &
  \textbf{Classification Accuracy} &
  78.12\% &
  56.25\% &
  75.00\% &
  71.88\% &
  59.38\% &
  65.62\% &
  62.50\% &
  50.00\% \\ \cline{1-2}
\multicolumn{1}{|c|}{\multirow{2}{*}{\textbf{VGG19}}} &
  \textbf{Highest Validation Accuracy} &
  80.00\% &
  80.00\% &
  60.00\% &
  \textbf{92.00\%} &
  \textbf{72.00\%} &
  64.00\% &
  76.00\% &
  52.00\% \\ \cline{2-2}
\multicolumn{1}{|c|}{} &
  \textbf{Classification Accuracy} &
  75.00\% &
  68.75\% &
  50.00\% &
  71.88\% &
  78.12\% &
  65.62\% &
  56.25\% &
  46.88\% \\ \cline{1-2}
\multicolumn{1}{|c|}{\multirow{2}{*}{\textbf{ResNet50}}} &
  \textbf{Highest Validation Accuracy} &
  84.00\% &
  \textbf{84.00\%} &
  80.00\% &
  88.00\% &
  60.00\% &
  \textbf{84.00\%} &
  76.00\% &
  60.00\% \\ \cline{2-2}
\multicolumn{1}{|c|}{} &
  \textbf{Classification Accuracy} &
  87.50\% &
  68.75\% &
  62.50\% &
  68.75\% &
  68.75\% &
  68.75\% &
  65.62\% &
  40.62\% \\ \cline{1-2}
\multicolumn{1}{|c|}{\multirow{2}{*}{\textbf{ResNet101}}} &
  \textbf{Highest Validation Accuracy} &
  80.00\% &
  80.00\% &
  \textbf{88.00\%} &
  \textbf{92.00\%} &
  64.00\% &
  76.00\% &
  \textbf{80.00\%} &
  52.00\% \\ \cline{2-2}
\multicolumn{1}{|c|}{} &
  \textbf{Classification Accuracy} &
  78.12\% &
  68.75\% &
  62.50\% &
  81.25\% &
  46.88\% &
  65.62\% &
  65.62\% &
  46.88\% \\ \cline{1-2}
\multicolumn{1}{|c|}{\multirow{2}{*}{\textbf{ResNet152}}} &
  \textbf{Highest Validation Accuracy} &
  \textbf{92.00\%} &
  68.00\% &
  60.00\% &
  \textbf{92.00\%} &
  68.00\% &
  72.00\% &
  72.00\% &
  52.00\% \\ \cline{2-2}
\multicolumn{1}{|c|}{} &
  \textbf{Classification Accuracy} &
  81.25\% &
  78.12\% &
  56.25\% &
  62.50\% &
  62.50\% &
  62.50\% &
  71.88\% &
  46.88\% \\ \cline{1-2}
\multicolumn{1}{|c|}{\multirow{2}{*}{\textbf{EfficientNetB0}}} &
  \textbf{Highest Validation Accuracy} &
  76.00\% &
  72.00\% &
  80.00\% &
  76.00\% &
  36.00\% &
  \textbf{84.00\%} &
  \textbf{80.00\%} &
  \textbf{64.00\%} \\ \cline{2-2}
\multicolumn{1}{|c|}{} &
  \textbf{Classification Accuracy} &
  75.00\% &
  84.38\% &
  62.50\% &
  78.12\% &
  62.50\% &
  65.62\% &
  65.62\% &
  53.12\% \\ \cline{1-2}
\multicolumn{1}{|c|}{\multirow{2}{*}{\textbf{EfficientNetB1}}} &
  \textbf{Highest Validation Accuracy} &
  76.00\% &
  68.00\% &
  60.00\% &
  88.00\% &
  40.00\% &
  80.00\% &
  76.00\% &
  56.00\% \\ \cline{2-2}
\multicolumn{1}{|c|}{} &
  \textbf{Classification Accuracy} &
  78.12\% &
  68.75\% &
  50.00\% &
  84.38\% &
  56.25\% &
  71.88\% &
  65.62\% &
  46.88\% \\ \cline{1-2}
\multicolumn{1}{|c|}{\multirow{2}{*}{\textbf{EfficientNetB2}}} &
  \textbf{Highest Validation Accuracy} &
  80.00\% &
  76.00\% &
  84.00\% &
  80.00\% &
  44.00\% &
  \textbf{84.00\%} &
  76.00\% &
  56.00\% \\ \cline{2-2}
\multicolumn{1}{|c|}{} &
  \textbf{Classification Accuracy} &
  78.12\% &
  81.25\% &
  75.00\% &
  81.25\% &
  59.38\% &
  68.75\% &
  62.50\% &
  43.75\% \\ \cline{1-2}
\multicolumn{1}{|c|}{\multirow{2}{*}{\textbf{EfficientNetB3}}} &
  \textbf{Highest Validation Accuracy} &
  \textbf{92.00\%} &
  64.00\% &
  80.00\% &
  \textbf{92.00\%} &
  48.00\% &
  \textbf{84.00\%} &
  76.00\% &
  52.00\% \\ \cline{2-2}
\multicolumn{1}{|c|}{} &
  \textbf{Classification Accuracy} &
  81.25\% &
  75.00\% &
  75.00\% &
  81.25\% &
  62.50\% &
  78.12\% &
  59.38\% &
  50.00\% \\ \cline{1-2}
\multicolumn{1}{|c|}{\multirow{2}{*}{\textbf{EfficientNetB4}}} &
  \textbf{Highest Validation Accuracy} &
  64.00\% &
  80.00\% &
  \textbf{88.00\%} &
  88.00\% &
  48.00\% &
  \textbf{84.00\%} &
  68.00\% &
  56.00\% \\ \cline{2-2}
\multicolumn{1}{|c|}{} &
  \textbf{Classification Accuracy} &
  62.50\% &
  78.12\% &
  75.00\% &
  71.88\% &
  56.25\% &
  71.88\% &
  71.88\% &
  65.62\% \\ 
  \hline
\end{tabular}%
}
\vspace{.1mm}
\end{table}

\begin{table}[H]
\centering
\caption{Results for Domestic class}
\label{tab:domestic-results}
\resizebox{\textwidth}{!}{%
\begin{tabular}{|cc|cccccccc|}
\hline
\multicolumn{2}{|c|}{\textbf{Filtration Mode}} &
  \multicolumn{1}{c|}{\multirow{2}{*}{\textbf{No Filter}}} &
  \multicolumn{1}{c|}{\multirow{2}{*}{\textbf{Noise Removal}}} &
  \multicolumn{1}{c|}{\multirow{2}{*}{\textbf{PCEN}}} &
  \multicolumn{1}{c|}{\multirow{2}{*}{\textbf{Audio Crop}}} &
  \multicolumn{1}{c|}{\multirow{2}{*}{\textbf{Low Pass Filter}}} &
  \multicolumn{1}{c|}{\multirow{2}{*}{\textbf{High Pass Filter}}} &
  \multicolumn{1}{c|}{\multirow{2}{*}{\textbf{Band Pass Filter}}} &
  \multicolumn{1}{c|}{\multirow{2}{*}{\textbf{Band Stop Filter}}} \\ \cline{1-2}
\multicolumn{1}{|c|}{\textbf{CNN Models}} &
  \textbf{Accuracies} &
  \multicolumn{1}{c|}{} &
  \multicolumn{1}{c|}{} &
  \multicolumn{1}{c|}{} &
  \multicolumn{1}{c|}{} &
  \multicolumn{1}{c|}{} &
  \multicolumn{1}{c|}{} &
  \multicolumn{1}{c|}{} &
  \multicolumn{1}{c|}{} \\ \hline
\multicolumn{1}{|c|}{\multirow{2}{*}{\textbf{VGG16}}} &
  \textbf{Highest Validation Accuracy} &
  \textbf{96.08\%} &
  94.12\% &
  50.98\% &
  86.27\% &
  60.78\% &
  72.55\% &
  70.59\% &
  45.10\% \\ \cline{2-2}
\multicolumn{1}{|c|}{} &
  \textbf{Classification Accuracy} &
  82.81\% &
  84.38\% &
  46.88\% &
  89.06\% &
  57.81\% &
  70.31\% &
  59.38\% &
  43.75\% \\ \cline{1-2}
\multicolumn{1}{|c|}{\multirow{2}{*}{\textbf{VGG19}}} &
  \textbf{Highest Validation Accuracy} &
  82.35\% &
  88.24\% &
  56.86\% &
  84.31\% &
  64.71\% &
  76.47\% &
  74.51\% &
  50.98\% \\ \cline{2-2}
\multicolumn{1}{|c|}{} &
  \textbf{Classification Accuracy} &
  82.81\% &
  90.62\% &
  60.94\% &
  87.50\% &
  62.50\% &
  65.62\% &
  65.62\% &
  37.50\% \\ \cline{1-2}
\multicolumn{1}{|c|}{\multirow{2}{*}{\textbf{ResNet50}}} &
  \textbf{Highest Validation Accuracy} &
  86.27\% &
  92.16\% &
  62.75\% &
  92.00\% &
  52.94\% &
  72.55\% &
  62.75\% &
  43.14\% \\ \cline{2-2}
\multicolumn{1}{|c|}{} &
  \textbf{Classification Accuracy} &
  78.12\% &
  92.19\% &
  60.94\% &
  94.12\% &
  65.62\% &
  70.31\% &
  73.44\% &
  43.75\% \\ \cline{1-2}
\multicolumn{1}{|c|}{\multirow{2}{*}{\textbf{ResNet101}}} &
  \textbf{Highest Validation Accuracy} &
  94.12\% &
  96.08\% &
  68.63\% &
  92.16\% &
  58.82\% &
  74.51\% &
  74.51\% &
  47.06\% \\ \cline{2-2}
\multicolumn{1}{|c|}{} &
  \textbf{Classification Accuracy} &
  87.50\% &
  90.62\% &
  71.88\% &
  87.50\% &
  68.75\% &
  68.75\% &
  71.88\% &
  39.06\% \\ \cline{1-2}
\multicolumn{1}{|c|}{\multirow{2}{*}{\textbf{ResNet152}}} &
  \textbf{Highest Validation Accuracy} &
  \textbf{96.08\%} &
  92.16\% &
  60.78\% &
  \textbf{96.08\%} &
  56.86\% &
  78.43\% &
  68.63\% &
  43.14\% \\ \cline{2-2}
\multicolumn{1}{|c|}{} &
  \textbf{Classification Accuracy} &
  87.50\% &
  89.06\% &
  54.69\% &
  87.50\% &
  65.62\% &
  65.62\% &
  64.06\% &
  34.38\% \\ \cline{1-2}
\multicolumn{1}{|c|}{\multirow{2}{*}{\textbf{EfficientNetB0}}} &
  \textbf{Highest Validation Accuracy} &
  \textbf{96.08\%} &
  94.12\% &
  72.55\% &
  94.12\% &
  \textbf{66.67\%} &
  68.63\% &
  82.35\% &
  \textbf{56.86\%} \\ \cline{2-2}
\multicolumn{1}{|c|}{} &
  \textbf{Classification Accuracy} &
  84.38\% &
  89.06\% &
  67.19\% &
  90.62\% &
  70.31\% &
  59.38\% &
  68.75\% &
  59.38\% \\ \cline{1-2}
\multicolumn{1}{|c|}{\multirow{2}{*}{\textbf{EfficientNetB1}}} &
  \textbf{Highest Validation Accuracy} &
  92.16\% &
  96.08\% &
  62.75\% &
  92.16\% &
  62.75\% &
  78.43\% &
  \textbf{86.27\%} &
  \textbf{56.86\%} \\ \cline{2-2}
\multicolumn{1}{|c|}{} &
  \textbf{Classification Accuracy} &
  89.06\% &
  87.50\% &
  48.44\% &
  92.19\% &
  70.31\% &
  62.50\% &
  70.31\% &
  50.00\% \\ \cline{1-2}
\multicolumn{1}{|c|}{\multirow{2}{*}{\textbf{EfficientNetB2}}} &
  \textbf{Highest Validation Accuracy} &
  94.12\% &
  96.08\% &
  \textbf{74.51\%} &
  94.12\% &
  62.75\% &
  76.47\% &
  84.31\% &
  52.94\% \\ \cline{2-2}
\multicolumn{1}{|c|}{} &
  \textbf{Classification Accuracy} &
  85.94\% &
  90.62\% &
  60.94\% &
  93.75\% &
  64.06\% &
  68.75\% &
  65.62\% &
  56.25\% \\ \cline{1-2}
\multicolumn{1}{|c|}{\multirow{2}{*}{\textbf{EfficientNetB3}}} &
  \textbf{Highest Validation Accuracy} &
  \textbf{96.08\%} &
  96.08\% &
  70.59\% &
  86.27\% &
  56.86\% &
  \textbf{82.35\%} &
  \textbf{86.27\%} &
  54.90\% \\ \cline{2-2}
\multicolumn{1}{|c|}{} &
  \textbf{Classification Accuracy} &
  89.06\% &
  90.62\% &
  60.94\% &
  85.94\% &
  64.06\% &
  73.44\% &
  73.44\% &
  56.25\% \\ \cline{1-2}
\multicolumn{1}{|c|}{\multirow{2}{*}{\textbf{EfficientNetB4}}} &
  \textbf{Highest Validation Accuracy} &
  \textbf{96.08\%} &
  \textbf{98.04\%} &
  52.94\% &
  88.24\% &
  50.98\% &
  \textbf{82.35\%} &
  82.35\% &
  43.14\% \\ \cline{2-2}
\multicolumn{1}{|c|}{} &
  \textbf{Classification Accuracy} &
  79.69\% &
  93.75\% &
  62.50\% &
  89.06\% &
  68.75\% &
  70.31\% &
  76.56\% &
  45.31\% \\ 
  \hline
\end{tabular}%
}
\vspace{.1mm}
\end{table}

\subsubsection{Domestic}
\label{subsubsec:domestic}

The results from table \ref{tab:domestic-results} of the Level 2 Classification of the Domestic class show that the highest validation accuracy as 98.04\% with Noise Removal and EfficientNetB1. It also showed the score of 96.08\% without application of modifiers and with Audio Crop. But, similar to the previous results of the Level 2 Classification, the lowest score was obtained as 56.86\% with Band Stop Filter.

\subsubsection{Outdoor}
\label{subsubsec:outdoor}

The Level 2 Classification of the Outdoor class got the highest validation accuracy as 93.75\% with the combination of Audio Crop and EfficientNetB3 as can be seen from table \ref{tab:human-results}. But it also showed the accuracy score high as 89.00\% and 85.94\%  with the application of Noise Removal and from the spectrograms of the raw audio files with VGG16 and ResNet152, respectively. The lowest validation accuracy obtaned in this case is 50.00\% with Band Stop Filter.

Now, comparing with the scores of previous works as shown in table \ref{tab:comparison}, we have achieved quite a challenging accuracy scores compared to the previous works.

\begin{table}[H]
\centering
\caption{Results for Outdoor class}
\label{tab:outdoor-results}
\resizebox{\textwidth}{!}{%
\begin{tabular}{|cc|cccccccc|}
\hline
\multicolumn{2}{|c|}{\textbf{Filtration Mode}} &
  \multicolumn{1}{c|}{\multirow{2}{*}{\textbf{No Filter}}} &
  \multicolumn{1}{c|}{\multirow{2}{*}{\textbf{Noise Removal}}} &
  \multicolumn{1}{c|}{\multirow{2}{*}{\textbf{PCEN}}} &
  \multicolumn{1}{c|}{\multirow{2}{*}{\textbf{Audio Crop}}} &
  \multicolumn{1}{c|}{\multirow{2}{*}{\textbf{Low Pass Filter}}} &
  \multicolumn{1}{c|}{\multirow{2}{*}{\textbf{High Pass Filter}}} &
  \multicolumn{1}{c|}{\multirow{2}{*}{\textbf{Band Pass Filter}}} &
  \multicolumn{1}{c|}{\multirow{2}{*}{\textbf{Band Stop Filter}}} \\ \cline{1-2}
\multicolumn{1}{|c|}{\textbf{CNN Models}} &
  \textbf{Accuracies} &
  \multicolumn{1}{c|}{} &
  \multicolumn{1}{c|}{} &
  \multicolumn{1}{c|}{} &
  \multicolumn{1}{c|}{} &
  \multicolumn{1}{c|}{} &
  \multicolumn{1}{c|}{} &
  \multicolumn{1}{c|}{} &
  \multicolumn{1}{c|}{} \\ \hline
\multicolumn{1}{|c|}{\multirow{2}{*}{\textbf{VGG16}}} &
  \textbf{Highest Validation Accuracy} &
  78.12\% &
  \textbf{89.06\%} &
  59.38\% &
  89.06\% &
  53.12\% &
  68.75\% &
  57.81\% &
  35.94\% \\ \cline{2-2}
\multicolumn{1}{|c|}{} &
  \textbf{Classification Accuracy} &
  78.75\% &
  82.50\% &
  48.75\% &
  76.25\% &
  53.75\% &
  47.50\% &
  46.25\% &
  36.25\% \\ \cline{1-2}
\multicolumn{1}{|c|}{\multirow{2}{*}{\textbf{VGG19}}} &
  \textbf{Highest Validation Accuracy} &
  79.69\% &
  76.56\% &
  60.94\% &
  84.38\% &
  60.94\% &
  60.94\% &
  60.94\% &
  40.62\% \\ \cline{2-2}
\multicolumn{1}{|c|}{} &
  \textbf{Classification Accuracy} &
  83.75\% &
  76.25\% &
  50.00\% &
  80.00\% &
  57.50\% &
  42.50\% &
  48.75\% &
  35.00\% \\ \cline{1-2}
\multicolumn{1}{|c|}{\multirow{2}{*}{\textbf{ResNet50}}} &
  \textbf{Highest Validation Accuracy} &
  82.81\% &
  79.69\% &
  64.06\% &
  87.50\% &
  57.81\% &
  75.00\% &
  62.50\% &
  46.88\% \\ \cline{2-2}
\multicolumn{1}{|c|}{} &
  \textbf{Classification Accuracy} &
  83.75\% &
  81.25\% &
  57.50\% &
  83.75\% &
  63.75\% &
  58.75\% &
  60.00\% &
  37.50\% \\ \cline{1-2}
\multicolumn{1}{|c|}{\multirow{2}{*}{\textbf{ResNet101}}} &
  \textbf{Highest Validation Accuracy} &
  81.25\% &
  75.00\% &
  59.38\% &
  79.69\% &
  57.81\% &
  \textbf{76.56\%} &
  65.62\% &
  43.75\% \\ \cline{2-2}
\multicolumn{1}{|c|}{} &
  \textbf{Classification Accuracy} &
  83.75\% &
  80.00\% &
  50.00\% &
  77.50\% &
  71.25\% &
  60.00\% &
  58.75\% &
  28.75\% \\ \cline{1-2}
\multicolumn{1}{|c|}{\multirow{2}{*}{\textbf{ResNet152}}} &
  \textbf{Highest Validation Accuracy} &
  \textbf{85.94\%} &
  78.12\% &
  56.25\% &
  87.50\% &
  48.44\% &
  71.88\% &
  57.81\% &
  43.75\% \\ \cline{2-2}
\multicolumn{1}{|c|}{} &
  \textbf{Classification Accuracy} &
  88.75\% &
  80.00\% &
  55.00\% &
  82.50\% &
  60.00\% &
  55.00\% &
  55.00\% &
  33.75\% \\ \cline{1-2}
\multicolumn{1}{|c|}{\multirow{2}{*}{\textbf{EfficientNetB0}}} &
  \textbf{Highest Validation Accuracy} &
  84.38\% &
  81.25\% &
  \textbf{70.31\%} &
  89.06\% &
  56.25\% &
  73.44\% &
  \textbf{68.75\%} &
  48.44\% \\ \cline{2-2}
\multicolumn{1}{|c|}{} &
  \textbf{Classification Accuracy} &
  85.00\% &
  85.00\% &
  60.00\% &
  80.00\% &
  65.00\% &
  53.75\% &
  55.00\% &
  36.25\% \\ \cline{1-2}
\multicolumn{1}{|c|}{\multirow{2}{*}{\textbf{EfficientNetB1}}} &
  \textbf{Highest Validation Accuracy} &
  81.25\% &
  75.00\% &
  60.94\% &
  89.06\% &
  59.38\% &
  75.00\% &
  \textbf{68.75\%} &
  \textbf{50.00\%} \\ \cline{2-2}
\multicolumn{1}{|c|}{} &
  \textbf{Classification Accuracy} &
  82.50\% &
  83.75\% &
  61.25\% &
  81.25\% &
  63.75\% &
  62.50\% &
  60.00\% &
  42.50\% \\ \cline{1-2}
\multicolumn{1}{|c|}{\multirow{2}{*}{\textbf{EfficientNetB2}}} &
  \textbf{Highest Validation Accuracy} &
  81.25\% &
  79.69\% &
  64.06\% &
  82.81\% &
  60.94\% &
  70.31\% &
  64.06\% &
  45.31\% \\ \cline{2-2}
\multicolumn{1}{|c|}{} &
  \textbf{Classification Accuracy} &
  87.50\% &
  85.00\% &
  53.75\% &
  77.50\% &
  63.75\% &
  58.75\% &
  55.00\% &
  43.75\% \\ \cline{1-2}
\multicolumn{1}{|c|}{\multirow{2}{*}{\textbf{EfficientNetB3}}} &
  \textbf{Highest Validation Accuracy} &
  84.38\% &
  85.94\% &
  67.19\% &
  \textbf{93.75\%} &
  \textbf{62.50\%} &
  \textbf{76.56\%} &
  \textbf{68.75\%} &
  43.75\% \\ \cline{2-2}
\multicolumn{1}{|c|}{} &
  \textbf{Classification Accuracy} &
  85.00\% &
  82.50\% &
  52.50\% &
  82.50\% &
  63.75\% &
  58.75\% &
  61.25\% &
  30.00\% \\ \cline{1-2}
\multicolumn{1}{|c|}{\multirow{2}{*}{\textbf{EfficientNetB4}}} &
  \textbf{Highest Validation Accuracy} &
  78.12\% &
  81.25\% &
  67.19\% &
  85.94\% &
  56.25\% &
  70.31\% &
  62.50\% &
  45.31\% \\ \cline{2-2}
\multicolumn{1}{|c|}{} &
  \textbf{Classification Accuracy} &
  82.50\% &
  81.25\% &
  56.25\% &
  82.50\% &
  56.25\% &
  65.00\% &
  58.75\% &
  40.00\% \\ 
  \hline
\end{tabular}%
}
\end{table}

\begin{table}[h]
\centering
\caption{Comparison with previous works}
\label{tab:comparison}
\begin{tabular}{|c|c|l}
\cline{1-2}
\textbf{Method by}                        & \textbf{Score on ESC-50}                                                                                                                           &  \\ \cline{1-2}
Piczak                                    & 64.50\%                                                                                                                                            &  \\ \cline{1-2}
Agrawal et al.                            & 81.95\%                                                                                                                                            &  \\ \cline{1-2}
Zhang et al.                              & 68.10\%                                                                                                                                            &  \\ \cline{1-2}
\multirow{2}{*}{Zhichao et al.}           & 83.90\%                                                                                                                                            &  \\ \cline{2-2}
                                          & 86.50\%                                                                                                                                            &  \\ \cline{1-2}
Ullo et al.                               & 95.80\%                                                                                                                                            &  \\ \cline{1-2}
Mushtaq et al.                            & 97.57\%                                                                                                                                            &  \\ \cline{1-2}
\multirow{2}{*}{Two-Level Classification} & \textbf{Level 1 } - 78.75\%                                                                                                                       &  \\ \cline{2-2}
                                          & \textbf{Level 2} - 
                                          98.04 - (Highest) \\

                                          \cline{1-2}
\end{tabular}%
\end{table}

\section{Discussion}
\label{discussion}

In this section, we are going to discuss the results obtained and shown in Section \ref{results}. As from table \ref{tab:level-1-classifier} it can be seen that the Level 1 Classification got its highest Classification Accuracy, 78.75\% from the CNN Model EfficientNetB2 with Audio Crop. 

\begin{figure}
\centering
\begin{subfigure}{.5\textwidth}
  \centering
  \includegraphics[width=\linewidth]{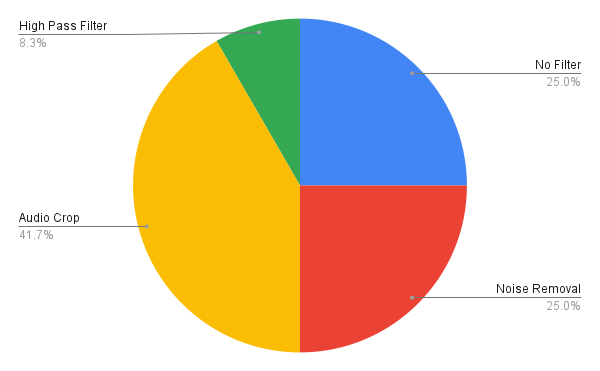}
  \caption{Distribution of the best Audio Modifiers}
  \label{fig:best-audio-modifiers}
\end{subfigure}%
\begin{subfigure}{.5\textwidth}
  \centering
  \includegraphics[width=\linewidth]{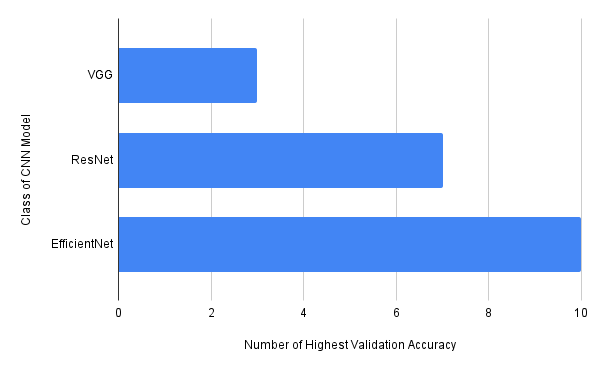}
  \caption{Visualization of the best scores obtained CNN classes}
  \label{fig:best-cnn-classes}
\end{subfigure}
\caption{Visualization of the number of best scores obtained by Audio Modifiers and CNN Models}
\end{figure}

Now in cases of the other classifications, the best results have been shown in the table \ref{tab:best-results}. Table \ref{tab:best-cnn-classes} and Fig \ref{fig:best-cnn-classes} shows the number of times each CNN class achieved the highest validation scores and it is quite clear from table \ref{tab:best-cnn-classes} and Fig \ref{fig:best-cnn-classes} that, EfficientNet worked best in most of the cases. Though it has got Highest Validation Accuracy in 10 cases but, if we examine the tables \ref{tab:level-1-classifier}, \ref{tab:animal-results}, \ref{tab:bird-results}, \ref{tab:nature-results}, \ref{tab:human-results}, \ref{tab:machine-sounds-results}, \ref{tab:domestic-results} and \ref{tab:outdoor-results} in Section \ref{results}, it will be clear that, in most of the cases the audio modifiers have got their highest Classification Accuracies, in case of Level 1 Classification and Highest Validation Accuracies in case of the other classifications with a CNN model belonging to the EfficientNet. 

Now coming to the other crucial part of the research, the audio modifier, Audio Crop gave the best scores in most of the cases as shown in table \ref{tab:best-audio-modifiers} and Fig \ref{fig:best-audio-modifiers}. As found in the table \ref{tab:best-results}, Audio Crop has got Highest Validation Accuracy for each of the Animal, Birds, Human, Machine Sounds and Outdoor classifications, but from the tables \ref{tab:nature-results} and \ref{tab:domestic-results}, Audio Crop might have failed to give the Highest Validation Accuracy, but it worked at par with the other audio modifiers.

\begin{table}[h]
\centering
\caption{Best Results obtained with CNN Models and Audio Modifiers
}
\label{tab:best-results}
\resizebox{\textwidth}{!}{%
\begin{tabular}{|c|c|c|c|}
\hline
\textbf{Mode of Classification}          & \textbf{CNN Model} & \textbf{Audio Modifier}    & \textbf{Highest Validation Accuracy} \\ \hline
\textbf{Animal}   & EfficientNetB2 & Audio Crop                     & 88.24\% \\ \hline
\multirow{6}{*}{\textbf{Birds}}          & VGG16              & \multirow{2}{*}{No Filter} & \multirow{6}{*}{96.00\%}             \\ \cline{2-2}
                  & EfficientNetB2 &                                &         \\ \cline{2-3}
                  & EfficientNetB1 & Noise Removal                  &         \\ \cline{2-3}
                  & ResNet152      & \multirow{2}{*}{Audio Crop}    &         \\ \cline{2-2}
                  & EfficientNetB1 &                                &         \\ \cline{2-3}
                  & ResNet50       & High Pass Filter               &         \\ \hline
\multirow{3}{*}{\textbf{Nature}}         & ResNet152          & No Filter                  & \multirow{3}{*}{95.45\%}             \\ \cline{2-3}
                  & ResNet152      & \multirow{2}{*}{Noise Removal} &         \\ \cline{2-2}
                  & EfficientNetB1 &                                &         \\ \hline
\textbf{Human}    & EfficientNetB0 & Audio Crop                     & 93.75\% \\ \hline
\multirow{7}{*}{\textbf{Machine Sounds}} & VGG16              & \multirow{3}{*}{No Filter} & \multirow{7}{*}{92.00\%}             \\ \cline{2-2}
                  & ResNet152      &                                &         \\ \cline{2-2}
                  & EfficientNetB3 &                                &         \\ \cline{2-3}
                  & VGG19          & \multirow{4}{*}{Audio Crop}    &         \\ \cline{2-2}
                  & ResNet101      &                                &         \\ \cline{2-2}
                  & ResNet152      &                                &         \\ \cline{2-2}
                  & EfficientNetB3 &                                &         \\ \hline
\textbf{Domestic} & EfficientNetB4 & Noise Removal                  & 98.04\% \\ \hline
\textbf{Outdoor}  & EfficientNetB3 & Audio Crop                     & 93.75\% \\ \hline
\end{tabular}%
}
\end{table}

\begin{table}[h!]
    \caption{Count of best scores by CNN Models and Audio Modifiers}

    \begin{subtable}[h]{\textwidth}

    \begin{center}

    \caption{Count of best scores by Audio Modifiers}
    \label{tab:best-audio-modifiers}
    \resizebox{\textwidth}{!}{%
    \begin{tabular}{|c|c|}
        \hline
        \textbf{Audio Modifier} & \textbf{Number of classification in which Highest Validation Accuracy is obtained} \\ \hline
        No Filter        & 3 \\ \hline
        Noise Removal    & 3 \\ \hline
        Audio Crop       & 5 \\ \hline
        High Pass Filter & 1 \\ \hline
    \end{tabular}%
    }

    \end{center}
    
    \end{subtable}
    
    \begin{subtable}[h]{\textwidth}
    \begin{center}
    \caption{Count of best scores by CNN Models}
    \label{tab:best-cnn-classes}
    \begin{tabular}{|c|c|}

        \hline
        \textbf{CNN Model Class} & \textbf{Number of times to get best scores} \\ \hline
        VGG                      & 3                                           \\ \hline
        ResNet                   & 7                                           \\ \hline
        EfficientNet             & 10                                          \\ \hline
        \end{tabular}%

                \end{center}
    \end{subtable}

\end{table}

Table \ref{tab:best-results} shows the accuracies for CNN models with No Filtration, Audio Crop, Noise Removal and High Pass Filter, but in some cases, other audio modifiers also worked well as the Low Pass Filter in Birds, High Pass Filter and Band Pass Filter in Nature, Band Pass Filter in Human and Band Pass Filter and PCEN in Machine Sounds. On the other hand, the four Audio Filters did not work well for the Outdoor class. Coming to the CNNs, ResNet and VGG showed high scores in the case of the classifications with a fewer number of samples, whereas, EfficientNet performed well with the problems with more classes and samples.

\section{Conclusion}
\label{conclusion}

The main objective of this paper is to propose a Two-Level Sound Classification method for the Environmental Sound Classification problem. Experiments on the ESC-50 dataset show that the classification accuracy of the Level 1 Classification was obtained as high as 78.75\%, while the highest validation score obtained by the Level 2 Classification is 98.04\%. In addition, this paper also shows the efficiencies of different CNN models combined with different audio modifiers and discussed their impact on the audio files. In future, we plan to optimize the hyperparameters more accurately and examine the performances of other threshold frequencies with the audio filters, though we have obtained the highest accuracy with audio crop and further improve the performance of the proposed methodology. We hope our method and process of thinking will encourage future researchers to implement them in their research.

\bibliographystyle{unsrt}  

\begin{thebibliography}{10}

\bibitem{smarthome}
Michel Vacher, Jean-Fran{\c{c}}ois Serignat, and Stephane Chaillol.
\newblock Sound classification in a smart room environment: an approach using gmm and hmm methods.
\newblock In {\em The 4th IEEE Conference on Speech Technology and Human-Computer Dialogue (SpeD 2007), Publishing House of the Romanian Academy (Bucharest)}, volume~1, pages 135--146, 2007.

\bibitem{security}
Regunathan Radhakrishnan, Ajay Divakaran, and A~Smaragdis.
\newblock Audio analysis for surveillance applications.
\newblock In {\em IEEE Workshop on Applications of Signal Processing to Audio and Acoustics, 2005.}, pages 158--161. IEEE, 2005.

\bibitem{machinehearing}
Richard~F Lyon.
\newblock Machine hearing: An emerging field [exploratory dsp].
\newblock {\em IEEE signal processing magazine}, 27(5):131--139, 2010.

\bibitem{piczak2015cnn}
Karol~J. Piczak.
\newblock Environmental sound classification with convolutional neural networks.
\newblock In {\em 2015 IEEE 25th International Workshop on Machine Learning for Signal Processing (MLSP)}, pages 1--6, 2015.

\bibitem{agrawal2017}
Dharmesh~M Agrawal, Hardik~B Sailor, Meet~H Soni, and Hemant~A Patil.
\newblock Novel teo-based gammatone features for environmental sound classification.
\newblock In {\em 2017 25th European Signal Processing Conference (EUSIPCO)}, pages 1809--1813. IEEE, 2017.

\bibitem{baritelli2008mfcc}
F.~Beritelli and R.~Grasso.
\newblock A pattern recognition system for environmental sound classification based on mfccs and neural networks.
\newblock In {\em 2008 2nd International Conference on Signal Processing and Communication Systems}, pages 1--4, 2008.

\bibitem{zhichao2018}
Zhichao Zhang, Shugong Xu, Shan Cao, and Shunqing Zhang.
\newblock Deep convolutional neural network with mixup for environmental sound classification.
\newblock In {\em Chinese conference on pattern recognition and computer vision (prcv)}, pages 356--367. Springer, 2018.

\bibitem{zhichao2019}
Zhichao Zhang, Shugong Xu, Shunqing Zhang, Tianhao Qiao, and Shan Cao.
\newblock Learning attentive representations for environmental sound classification.
\newblock {\em IEEE Access}, 7:130327--130339, 2019.

\bibitem{mushtaq2021}
Zohaib Mushtaq, Shun-Feng Su, and Quoc-Viet Tran.
\newblock Spectral images based environmental sound classification using cnn with meaningful data augmentation.
\newblock {\em Applied Acoustics}, 172:107581, 2021.

\bibitem{xiaohu2017}
Xiaohu Zhang, Yuexian Zou, and Wei Shi.
\newblock Dilated convolution neural network with leakyrelu for environmental sound classification.
\newblock In {\em 2017 22nd international conference on digital signal processing (DSP)}, pages 1--5. IEEE, 2017.

\bibitem{ullo2020}
Silvia~Liberata Ullo, Smith~K Khare, Varun Bajaj, and GR~Sinha.
\newblock Hybrid computerized method for environmental sound classification.
\newblock {\em IEEE Access}, 8:124055--124065, 2020.

\bibitem{ansar2024efficientnet}
Wazib Ansar, Ahan Chatterjee, Saptarsi Goswami, and Amlan Chakrabarti.
\newblock An efficientnet-based ensemble for bird-call recognition with enhanced noise reduction.
\newblock {\em SN Computer Science}, 5(2):265, 2024.

\bibitem{lecunn1990}
Yann Lecun, B.~Boser, {J. S.} Denker, D.~Henderson, {R. E.} Howard, W.~Hubbard, and L.D. Jackel.
\newblock Handwritten digit recognition with a back-propagation network.
\newblock In {\em Advances in Neural Information Processing Systems (NIPS 1989), Denver, CO}, volume~2. Morgan Kaufmann, 1990.

\bibitem{krizhevsky2012}
Alex Krizhevsky, Ilya Sutskever, and Geoffrey~E Hinton.
\newblock Imagenet classification with deep convolutional neural networks.
\newblock {\em Advances in neural information processing systems}, 25, 2012.

\bibitem{vgg16}
Karen Simonyan and Andrew Zisserman.
\newblock Very deep convolutional networks for large-scale image recognition.
\newblock {\em arXiv preprint arXiv:1409.1556}, 2014.

\bibitem{resnet}
Kaiming He, Xiangyu Zhang, Shaoqing Ren, and Jian Sun.
\newblock Deep residual learning for image recognition.
\newblock In {\em Proceedings of the IEEE conference on computer vision and pattern recognition}, pages 770--778, 2016.

\bibitem{efficientnet}
Mingxing Tan and Quoc Le.
\newblock Efficientnet: Rethinking model scaling for convolutional neural networks.
\newblock In {\em International conference on machine learning}, pages 6105--6114. PMLR, 2019.

\bibitem{spectralgating}
Joshua~M Inouye, Silvia~S Blemker, and David~I Inouye.
\newblock Towards undistorted and noise-free speech in an mri scanner: correlation subtraction followed by spectral noise gating.
\newblock {\em The Journal of the Acoustical Society of America}, 135(3):1019--1022, 2014.

\bibitem{pcen1}
Yuxuan Wang, Pascal Getreuer, Thad Hughes, Richard~F Lyon, and Rif~A Saurous.
\newblock Trainable frontend for robust and far-field keyword spotting.
\newblock In {\em 2017 IEEE International Conference on Acoustics, Speech and Signal Processing (ICASSP)}, pages 5670--5674. IEEE, 2017.

\bibitem{pcen2}
Vincent Lostanlen, Justin Salamon, Mark Cartwright, Brian McFee, Andrew Farnsworth, Steve Kelling, and Juan~Pablo Bello.
\newblock Per-channel energy normalization: Why and how.
\newblock {\em IEEE Signal Processing Letters}, 26(1):39--43, 2018.

\bibitem{audiofilters}
RR~Porle, NS~Ruslan, NM~Ghani, NA~Arif, SR~Ismail, N~Parimon, and M~Mamat.
\newblock A survey of filter design for audio noise reduction.
\newblock {\em J. Adv. Rev. Sci. Res}, 12(1):26--44, 2015.

\bibitem{piczak2015data}
Karol~J Piczak.
\newblock Esc: Dataset for environmental sound classification.
\newblock In {\em Proceedings of the 23rd ACM international conference on Multimedia}, pages 1015--1018, 2015.

\end{thebibliography}

\end{document}